\documentclass[aps,prb,twocolumn,notitlepage,showpacs,superscriptaddress,10pt]{revtex4-2}
\usepackage{graphicx}
\usepackage{dcolumn}
\usepackage{bm}
\usepackage[colorlinks, allcolors=blue]{hyperref}
\usepackage[usenames, dvipsnames]{color}
\usepackage{float}

\usepackage{xcolor}
\usepackage{tabularx}
\usepackage{amsmath}
\usepackage{afterpage}
\usepackage{booktabs}
\usepackage{array}
\usepackage{setspace}
\graphicspath{{Figs/}}
\usepackage{hhline}
\usepackage{xfrac}
\usepackage{float,graphicx}
\usepackage{mathtools}
\usepackage{listings}
\usepackage{amssymb}
\usepackage{amsfonts}

\usepackage{epstopdf}
\catcode`@11
\def\seceqaa{\@addtoreset{equation}{section}
\def\theequation{A\arabic{equation}}}
\def\seceqbb{\@addtoreset{equation}{section}
\def\theequation{B\arabic{equation}}}
\def\seceqcc{\@addtoreset{equation}{section}
\def\theequation{C\arabic{equation}}}
\def\seceqdd{\@addtoreset{equation}{section}
\def\theequation{D\arabic{equation}}}
\def\seceqee{\@addtoreset{equation}{section}
\def\theequation{E\arabic{equation}}}
\def\seceqff{\@addtoreset{equation}{section}
\def\theequation{F\arabic{equation}}}
\def\seceqgg{\@addtoreset{equation}{section}
\def\theequation{G\arabic{equation}}}
\def\seceqhh{\@addtoreset{equation}{section}
\def\theequation{H\arabic{equation}}}
\catcode`@11

\newcommand{\up}{\uparrow}
\newcommand{\dn}{\downarrow}

\newcommand{\ket}[1]{|#1\rangle}

\begin{document}

\title{ Phase diagram of a coupled trimer system at half filling using the Hubbard model 
} 

\author{Sourabh Saha}
\affiliation{Department of Condensed Matter and Materials Physics,
S. N. Bose National Centre for Basic Sciences, JD Block, Sector III, Salt Lake, Kolkata 700106, India}
\author{Hosho Katsura}
\email{katsura@phys.s.u-tokyo.ac.jp}
\affiliation{Department of Physics, The University of Tokyo, 
Hongo, Bunkyo-ku, Tokyo 113-0033, Japan} 
\affiliation{Institute for Physics of Intelligence, The University of Tokyo, 
Hongo, Bunkyo-ku, Tokyo 113-0033, Japan} 
\affiliation{Trans-scale Quantum Science Institute, The University of Tokyo, Hongo, Bunkyo-ku, Tokyo 113-0033, Japan}
\author{Manoranjan Kumar}
\email{manoranjan.kumar@bose.res.in}
\thanks{Last two authors contributed equally.}
\affiliation{Department of Condensed Matter and Materials Physics,
S. N. Bose National Centre for Basic Sciences, JD Block, Sector III, Salt Lake, Kolkata 700106, India}
\date{\today}

\begin{abstract}

 Flat band systems have recently attracted significant attention due to their instability under small perturbations, which can lead to the stabilization of many exotic quantum phases. 
 We study a trimer ladder which shows a middle flat band in the absence of onsite Coulomb interaction. We investigate the quantum phases of the Hubbard model on this geometry using exact diagonalization (ED), density matrix renormalization group (DMRG), and perturbation theory. We construct a quantum phase diagram in the plane of the next-nearest-neighbor hopping parameter $t_2$ and onsite Coulomb interaction $U$, revealing five distinct quantum phases. At low $U$ and moderate to high magnitude of $t_2$, the system exhibits metallic behavior, while at large $U$ and small magnitude of $t_2$, it transitions to a ferrimagnetic insulator phase, similar to those observed in certain trimer materials. In the small $t_2$ limit, the Fermi energy is in the flat band, leading to localization of the electrons within the trimer. At low $U$ and small magnitude of $t_2$, the flat band mechanism favors insulating ferrimagnetism, whereas at large $U$, ferrimagnetic states emerge from singlet dimer formation between neighboring sites of a trimer and an isolated corner spin, which connect ferromagnetically. The insulating cell spin density wave
 phase displays an up-up-down-down spin configuration due to competing nearest neighbor hopping, $t_1$. Interestingly, in moderate $U$ and $|t_2|>0.3$, the ground state behaves like metallic Tomonaga-Luttinger liquid.

\end{abstract}

\maketitle
\section{INTRODUCTION}
Low-dimensional correlated fermionic systems offer a rich playground for investigating the interplay between band structure, Coulomb interaction, and quantum fluctuations due to confinement of electrons \cite{fazekas1999lecture}. For instance, a single-band Hubbard model in one dimension away from half-filling exhibits the typical characteristics of a Tomonaga-Luttinger liquid (TLL) state, featuring gapless spin and charge modes arising under repulsive interactions \cite{haldane1981luttinger,doi:10.1143/JPSJ.69.1000,schulz1990correlation,giamarchi2003quantum}. This unique characteristic allows low-energy excitations in TLLs to be effectively described using bosonic fields \cite{tomonaga1950remarks,luttinger1963exactly}, resulting in various intriguing physical phenomena, such as power-law correlations and critical behavior. 

In addition, some of these systems can host flat bands, where electrons are localized due to the destructive interference between hopping paths associated with lattice symmetries \cite{chen2024isolated,xian2021engineering,leykam2018artificial}, and such systems exhibit highly degenerate energy levels, leading to an enhanced density of states and pronounced electronic correlation effects. This unique scenario of flat band gives rise to instabilities that lead to correlated phases, such as superconductivity\cite{aoki2020theoretical,kobayashi2016superconductivity,mahyaeh2022superconducting,shahbazi2023revival}, ferromagnetic (FM) \cite{hase2023flat,pons2020flat}, and charge density waves phase \cite{hofmann2023superconductivity,teng2023magnetism} by applying small perturbations like pressure, temperature, and magnetic field.  

In recent years, many flat band systems have been found in nature, for example, structures like the kagome realized in CoSn \cite{kang2020topological}, pyrochlore realized in Sn${}_2$Nb${}_2$O${}_7$ \cite{hase2018possibility}, Lieb-lattice in Ba${}_2$CuO${}_{3+\delta}$ \cite{yamazaki2020superconducting}. Quasi-one-dimensional systems like diamond chain realized in Cu${}_3$(CO${}_3$)${}_2$(OH)${}_2$ \cite{10.1143/PTPS.159.1}, sawtooth chain realized in Fe${}_{10}$Gd${}_{10}$ \cite{baniodeh2018high}, stub lattices in photonic systems \cite{real2017flat}, artificially synthesized polymer chains of Lieb ladder, skip Creutz ladder realized in ultra-cold atoms, and many other systems \cite{neves2024crystal}, which host flat bands.

The magnetic properties of the ground state (gs) in fermionic models are quite interesting and depend on the geometry of the system and the nature of correlations.  In the non-interacting limit, generally electrons exhibit Pauli paramagnetism \cite{schumacher1956electron,schumacher1963paramagnetic,schultz1967observation,kaeck1968electron,lee2013pauli}.  However, in the presence of finite onsite Coulomb interaction, in the mean field limit, the FM phase can be explained using the Stoner criterion, which requires $UD(E_F) \ge 1$, where $U$ and $D(E_F)$ are the onsite Coulomb interaction term and the density of state at the Fermi energy, respectively \cite{stoner1938collective}.  Another mechanism for achieving an FM gs was introduced by Mielke, specifically for flat-band systems based on line graphs  \cite{AMielke_1991,AMielke_1991_1,AMielke_1992} and then by Tasaki \cite{tasaki1992ferromagnetism,tasaki1994stability,TasakiBook}.  Mielke-Tasaki (MT) mechanism requires only an infinitesimal $U$ to stabilize the FM gs. If the lowest band is nearly flat, the ferromagnetic state can still persist for sufficiently large Coulomb repulsion \cite{tasaki1995ferromagnetism,tasaki2003ferromagnetism,tamura2019ferromagnetism,tamura}. 
\textcolor{black}{More recently, a new mechanism leading to kinetic-energy-driven ferromagnetism has been proposed \cite{derzhko2014dispersion,muller2016hubbard}.}
However, there is no general recipe to achieve a ferrimagnetic gs of the Hubbard model on ladder systems. The only exception is Lieb's theorem, which applies exclusively to the model on a bipartite lattice at half-filling \cite{lieb1989two}.

\begin{figure}
\centering
\includegraphics[width=1.00\linewidth]{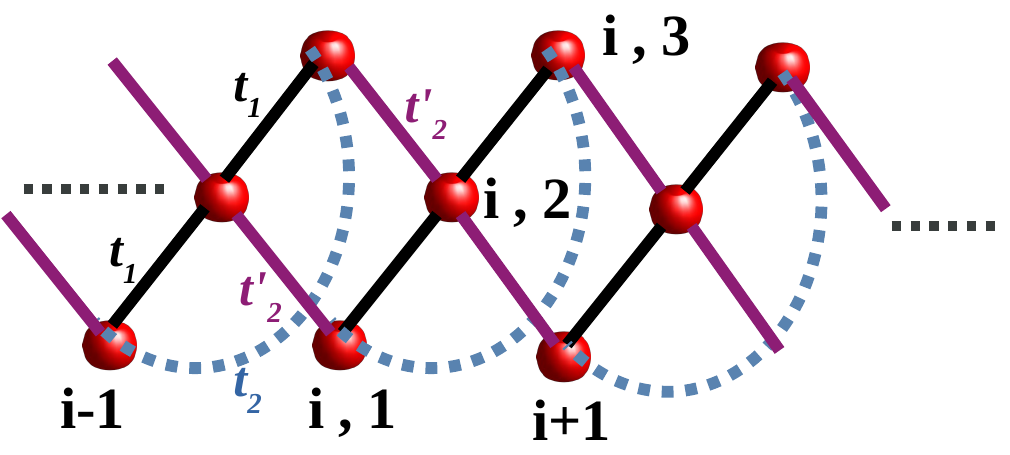}
\caption{Schematic representation of the lattice structure of coupled trimer system. Here, $t_1$ and $t_2$ are the strengths of the nearest-neighbor and next-nearest-neighbor hoppings within the same trimer, respectively, while $t'_2$ represents the strength of hopping between different trimer unit cells.
}
\label{fig1}
\end{figure}

In this paper, we investigate the quantum phase transitions in the Hubbard model on an interacting trimer ladder, depicted in Fig.\ref{fig1}. In large $U$ limit some of our results are applicable to the spin-$1/2$ system on distorted azurite materials like  Cu${}_3$(CO${}_3$)${}_2$(OH)${}_2$ \cite{ExperimentalObservationofCumaterial} and weakly interacting trimers in  Na${}_2$Cu${}_3$Ge${}_4$O${}_{12}$ \cite{bera2022emergent,yasui2014magnetic}. The organic compound PNNBNO forming an $AB_{2}$ structure also exhibits ferrimagnetism at low temperatures \cite{hosokoshi2001approach,yao2004transfer,yao2005thermodynamics}. The magnetic properties of these materials are modeled using the Heisenberg model \cite{montenegro2022ground}.

The trimer ladder can be mapped to a 3/4 skewed ladder \cite{giri2017quantum} and diamond lattice \cite{shahbazi2023revival} with next nearest neighbor as shown in Fig. S1 of the Supplemental Material.  In the strong Coulomb repulsion limit, i.e. $U/t_1 \gg 1$, where $t_1$ is the nearest neighbor hopping strength, the Hubbard model can be mapped to the spin-$1/2$ antiferromagnetic Heisenberg model on this system. The gs properties of the spin-$1/2$ antiferromagnetic $J_1$-$J_2$ Heisenberg model on this ladder have been studied, and the gs is known to vary from non-magnetic to ferrimagnetic by tuning the $J_2/J_1$ \cite{giri2017quantum}.  The ferrimagnetic ground state can be approximately described as a product of singlet between nearest neighbor spins and an isolated spin located at the other corner of the unit cell \cite{giri2017quantum}. However, a systematic study of the gs properties of the half-filled Hubbard model on a trimer ladder is absent in the literature. 

 In this work, we construct a quantum phase diagram of the Hubbard model on a trimer ladder in the parameter space defined by the nearest-neighbor hopping $t_1$, the next-nearest-neighbor hopping $t_2$, and the onsite Coulomb repulsion $U$. We show that there are five types of gs quantum phases: ferrimagnet,
 insulating cell spin density wave (ICSDW), metallic Tomonaga-Luttinger liquid I (MTLL I), metallic Tomonaga-Luttinger liquid II (MTLL II), and variable spin magnetic insulator (VSMI). These phases emerge by tuning the hopping parameter $t_2$ and the on-site interaction $U$. When $U$ is finite and magnitude of $t_2$ is sufficiently small, the gs exhibits a ferrimagnetic phase with an effective spin of one-third of the total spin. In the small $U$ limit, this ferrimagnetic ground state can be qualitatively explained by the presence of the nearly flat middle band, as we discuss below. On the other hand, in the large $U$ limit, the ferrimagnetism in the ground state can be understood through second-order perturbation theory applied to two neighboring trimers.
 This system also hosts Tomonaga-Luttinger liquid phases in the metallic gs. In the phase diagram, the ICSDW phase is sandwiched between the ferrimagnet
 and MTLL I phases.  At very high $U$ and moderate $t_2$,  
 VSMI phase emerges.  

The remainder of the paper is organized into four sections.  Section II discusses the model Hamiltonian and numerical methods. Results are discussed in Sec. III, which is divided into five subsections, which discuss the interacting and non-interacting limits of isolated and coupled trimer system. Different phases and their phase boundaries are also explained in this section. In  Sec. IV, we summarize and discuss the results.

\section{\label{sec2} MODEL AND METHODS}

We consider a repulsive single-band Hubbard model on a trimer ladder geometry shown in Fig. \ref{fig1}. The $i$th trimer (unit cell) consists of the three sites labeled $(i,1)$, $(i,2)$, and $(i,3)$, as shown in Fig. \ref{fig1}. For each site $(i,\alpha)$, we denote by $c^\dagger_{i,\alpha,\sigma}$ and $c_{i,\alpha,\sigma}$ the creation and annihilation operators, respectively, of an electron at $(i,\alpha)$ with spin $\sigma=\up,\dn$. The electron number operator is defined as $n_{i,\alpha,\sigma}=c^\dagger_{i,\alpha,\sigma}c_{i,\alpha,\sigma}$.

The whole Hamiltonian of the system can be divided into two parts: one is the sum of intra-trimer Hamiltonians, $H_{\rm trimer}^i$ for the $i$th trimer, and the other is the inter-trimer interaction $H_{\rm int}$. The model Hamiltonian for the trimer ladder can be written as
\begin{eqnarray}
     H =\sum_{i} H_{\rm trimer}^{i} + H_{\rm int},
    \label{eq1}
\end{eqnarray}
where
\begin{align}
& H_{\rm trimer}^{i}=t_{1}\sum_{\sigma=\up,\dn}(c^{\dagger}_{i,1,\sigma}c_{i,2,\sigma}+c^{\dagger}_{i,2,\sigma}c_{i,3,\sigma}+{\rm H.c.}) \nonumber\\
& +t_2\sum_{\sigma=\up,\dn}(c^{\dagger}_{i,1,\sigma}c_{i,3,\sigma}+{\rm H.c.})+U\sum_{\alpha=1}^{3}n_{i,\alpha,\uparrow}n_{i,\alpha,\downarrow},
\label{eq2x}\\
& H_{\rm int}=t_{2}' \sum_{i} \sum_{\sigma=\up,\dn}(c^{\dagger}_{i-1,3,\sigma}c_{i,2,\sigma}+c^{\dagger}_{i,2,\sigma}c_{i+1,1,\sigma} +{\rm H.c.}).
\label{eq2}
\end{align}

Here, $t_1$ and $t_2$, respectively, are the hopping parameters between nearest-neighbor and next-nearest-neighbor sites within the same trimer, while $t_2'$ represents the hopping parameter between different trimer unit cells. 
For studies of connected trimers, we have set
the intra and inter-trimer hopping parameter, $t_{2}$ and  $t_2'$ equal.
$U$ is the on-site Coulomb repulsive energy. We set $t_1=-1$ as the energy scale for our calculation and $t_2=t_2'\le0$.

In our study, we use exact diagonalization (ED) for small system sizes up to $N=12$ sites and the density matrix renormalization group (DMRG) method for larger systems up to $N=120$. The DMRG is a state-of-art numerical technique for solving many-body Hamiltonians in low-dimensional systems \cite{white1992density,schollwock2005density,hallberg2006new}. The accuracy of the results depends on the number of eigenvectors corresponding to the largest eigenvalues, $m$,
retained for the renormalization of operators and the Hamiltonian matrix of the system. We use open boundary conditions for the trimer system, keeping $m$ up to 500 and performing up to $10$ finite DMRG sweeps.

The Hamiltonian in Eq. (\ref{eq1}) conserves the total spin and its $z$ component, $S^z$, as well as the total electron number $N_{\rm e}=\sum_{i,\alpha,\sigma} n_{i,\alpha,\sigma}$. Therefore, the DMRG method is used to evaluate the gs and low-lying excited state properties in conserved $S^z$ sectors at fillings where the number of electrons $N_{\rm e}$ is conserved and equal to $N$.

\section{\label{sec1}RESULTS}
In this section, we will discuss the gs properties of the  Hamiltonian in Eq. (\ref{eq1}) on the ladder geometry and its basic unit, trimer in various $t_2$ limits. The main goal of this paper is to understand the gs phases in the $t_2$-$U$ parameter regime and to construct the quantum phase diagram of this model based on the spin gap, charge gap, and charge and spin correlations. We have also used the gs charge and spin densities and expectation value of hopping terms for further analysis of the phase diagram. We first discuss the results of the non-interacting electron or tight-binding limit for a single trimer and coupled trimers. Later, we will discuss the effect of the repulsive Coulomb interaction, $U$, on the isolated trimers and then on the coupled trimers. We construct the quantum phase diagram of the trimer ladder and examine the various phases along with their boundaries.

\subsection{Non-interacting limit of single trimer}
\label{sec:single_trimer_energies}
\begin{figure}[hbt!]
\centering
\includegraphics[width=1.00\linewidth]{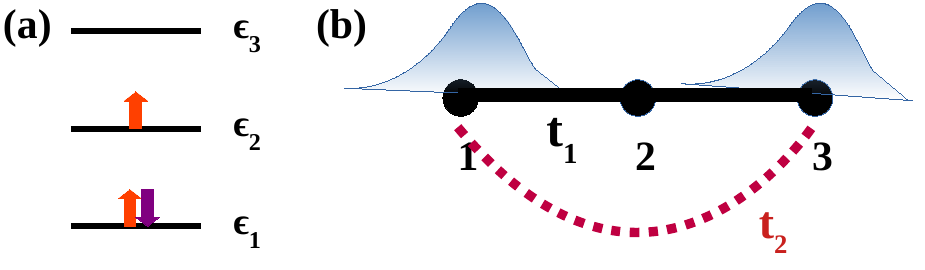}
\caption{(a) Energy levels of a trimer and its electron distribution at half filling in the non-interacting limit for 
$t_1<t_2\le 0$. (b) A schematic picture of the wave function ($\ket{\psi_2}$) distribution for the itinerant electron of a trimer along three different sites for 
$t_1<t_2\le 0$.}
\label{fig2}
\end{figure}

We begin by studying the tight-binding (TB) model for an isolated trimer, which is represented by the first two terms of the Hamiltonian in Eq. (\ref{eq2x}). In this equation, the sum over $i$ corresponds to all the trimers in the system. For simplicity, we focus on a single trimer and drop the index $i$ for the remainder of this subsection. Each isolated trimer, at half-filling, consists of three sites and three electrons. 
The system features two types of hopping: nearest-neighbor hopping $t_1$ and next-nearest-neighbor hopping $t_2$, as illustrated in Fig. \ref{fig2}(b). Depending on the value of $t_2$, the system can be visualized as either a triangular or a chain-like structure, with $t_2$ being finite or $t_2=0$, respectively. The TB Hamiltonian for the trimer is given by
\begin{align}
H_{\rm trimer}=\sum_{\sigma}[t_1(c^{\dagger}_{1,\sigma}c_{2,\sigma}+c^{\dagger}_{2,\sigma}c_{3,\sigma})+t_{2}c^{\dagger}_{1,\sigma}c_{3,\sigma}+{\rm H.c.}]. 
\label{eq3}
\end{align}

In a single-electron picture with fixed spin, one can think of $H_{\rm trimer}$ as a $3\times 3$ matrix. Diagonalizing this Hamiltonian matrix, we find the following eigenvalues and corresponding eigenvectors: 
\begin{eqnarray}
\begin{array}{ll}
\underline{\rm Eigenvalue} & \underline{\rm Eigenvector} \\
\epsilon_1 = \frac{1}{2} ( t_2 - \sqrt{8t^2_1+t^2_2} )~~ & \ket{\psi_1}= {\cal N}_1 ( \ket{1}-\frac{\epsilon_3}{t_1}\ket{2}+\ket{3} )\\
\epsilon_2 = -t_2~~ & \ket{\psi_2}= \frac{1}{\sqrt 2} (\ket{1}-\ket{3})\\
\epsilon_3 = \frac{1}{2}( t_2 + \sqrt{8t^2_1+t^2_2} )~~& \ket{\psi_3}= {\cal N}_3 ( \ket{1}-\frac{\epsilon_1}{t_1}\ket{2}+\ket{3} )
\end{array},\nonumber\\
\end{eqnarray}
where ${\cal N}_1$ and ${\cal N}_3$ are the normalization factors and  $\ket{i}$ is the local atomic orbital at site $i$. The gs electronic configuration for small $|t_2|$ is shown in Fig. \ref{fig2}(a). In this parameter regime, $\epsilon_1$ is doubly occupied and $\epsilon_2$ is singly occupied.  The state $\ket{\psi_1}$ is a linear combination of all the three orbitals $\ket{i}$ ($i=1,2,3$), whereas $\ket{\psi_2}$ has only contribution from the end site orbitals, as shown in Fig. \ref{fig2}(b).

For $t_2 = t_1$, $\epsilon_2$ and $\epsilon_3$ are degenerate, resulting in the ground state being four-fold degenerate. This occurs because the lowest energy state, $\epsilon_1$ is doubly occupied, whereas the singly occupied orbital is doubly degenerate and it can arrange in two possible ways. Another two-fold degeneracy comes from the spin degrees of freedom. Thus, the effective spin of the system is $1/2$, but there are two ways by which these electrons can be arranged, leading to two orbital degrees of freedom. In the case with $t_2<t_1$,  $\epsilon_1$ is the lowest energy state and the two electrons occupy this orbital, whereas $\epsilon_3$ is lower in energy than $\epsilon_2$. As a result, the orbital $\epsilon_3$ is singly occupied in the ground state. 
\begin{figure*}[t]
\begin{center}
\includegraphics[width=0.8\linewidth]{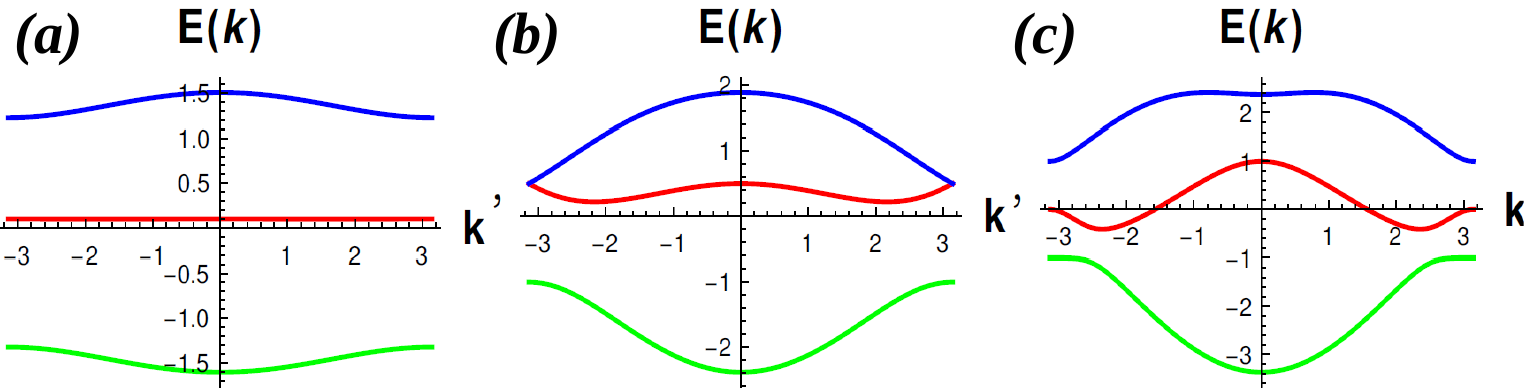}
\caption{(a),(b), and (c) Dispersion curves for three different  
values of $t_2=-0.1$, $-0.5$, and $-1.0$, respectively. For $t_2=-0.1$, the middle band is nearly flat,  
whereas for $t_2=-0.5$ and $-1.0$, it becomes dispersive.}
\label{fig3}
\end{center}
\end{figure*}

\subsection{Free electron on a connected trimer }
In this section, we consider the connected trimers with intra-trimer hopping term $t'_{2}=t_{2}$ as shown in Fig. \ref{fig1}. For free electrons, we set $U=0$ and the model in Eq. (\ref{eq1}) reduced to the TB model Hamiltonian given by
\begin{align}
H_{\rm TB} &= t_{1}\sum_{i,\sigma}(c^{\dagger}_{i,1,\sigma}c_{i,2,\sigma}+c^{\dagger}_{i,2,\sigma}c_{i,3,\sigma}+{\rm H.c.})\hspace{0.1cm}\nonumber\\ 
 &+t_2 \sum_{i,\sigma} (c^{\dagger}_{i,1,\sigma}c_{i,3,\sigma}+c^{\dagger}_{i-1,3,\sigma}c_{i,2,\sigma}\nonumber\\
 &+c^{\dagger}_{i,2,\sigma}c_{i+1,1,\sigma}\hspace{0.1cm}+{\rm H.c.}),\hspace{0.1cm}
\label{eq4}
\end{align}
 The Fourier transform of  $H_{\rm TB}$ can be expressed as a $3 \times 3$ matrix since each unit cell contains three lattice sites and forms a three-band model. Thus, the TB model Hamiltonian in $k$-space can be written as
\begin{align}
    H_{\rm TB} = \sum_{k,\sigma} \Psi^\dagger_{k,\sigma} {\sf H}(k) \Psi_{k,\sigma},
\end{align}
where $\Psi^{\dagger}_{k,\sigma}=(c^\dagger_{k,1,\sigma},c^\dagger_{k,2,\sigma},c^\dagger_{k,3,\sigma})$ and the momentum space Hamiltonian is given by
\begin{align}
    {\sf H}(k) = \begin{bmatrix}
    0  & t_{1}+t_{2}e^{-ik}  &  t_{2}\\
    t_{1}+t_{2}e^{ik}  &  0  & t_{1}+t_{2}e^{-ik}\\
    t_{2} &  t_{1}+t_{2}e^{ik}  & 0
    \end{bmatrix}.
\end{align}

The electronic dispersion curves, $E(k)$, of the three bands depend on the values of $t_2$ as shown in Fig. \ref{fig3}. It is interesting to note that the upper and lower bands are dispersive, whereas the middle band is nearly flat for $t_2> -0.5$; however, for $t_2 \leq-0.5$, the middle band has a higher propensity to disperse. The dispersion curves for three different values of $t_2=-0.1$, $-0.5$, and $-1.0$ are shown in Fig. \ref{fig3}(a), (b), and (c), respectively. For $t_2 \le -0.5$, the middle band has two minima at point $k= \pm 2\pi/3$ and maxima at $k=0$, which is more prominent for higher values of $|t_2|$.  In fact, the upper band exhibits the flat behavior around $k=0$ for large values of  $|t_2|$ as shown in Fig. \ref{fig3}(c). 

\begin{figure}[hbt!]
\centering
\includegraphics[width=1.00\linewidth]{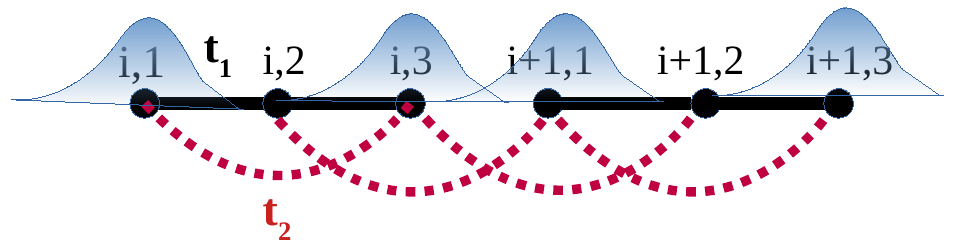}
\caption{Connected trimers and their wave function distribution for small $|t_2|$ limit. Here electrons are delocalized through the dotted bond, $t_2$.}
\label{fig4}
\end{figure}

  For small $|t_{2}|$, all three bands of the connected trimers originate from the three energy levels of the isolated trimer discussed in Sec. \ref{sec:single_trimer_energies}. The nearly flat band in this system arises due to the localized mid state of the isolated trimer. The upper and lower states of the isolated trimer are linear combinations of all three atomic orbital wave functions, whereas the mid state is formed by the linear combination of end-site orbitals only.  A pictorial representation of the flat band formation due to the localized wave function is shown in Fig. \ref{fig4}. Here, due to the intra-trimer hopping $t_2$, the wave functions are delocalized from the end sites, and the mid band becomes nearly flat, rather than fully flat for small $|t_2|$. The Hamiltonian in Eq. (\ref{eq4}) allows the hopping from the mid site of one trimer to the edge sites of another trimer, where each trimer has connections between its two edge sites. Therefore, the contribution of the first-order perturbation of $t_2$ in band formation is zero, and only the second-order perturbation of $t_2$ contributes to the mid-band dispersion. However, the lower and upper bands have contributions from the first-order and higher-order perturbations. The middle band becomes more dispersive in the case of large $t_2$, as shown in Fig. \ref{fig3}(b) and (c).  
 
  Another way to understand the nearly flat band for small $|t_{2}|$ is to note that the presence of sublattice (or chiral) symmetry when frustrated intra-trimer hopping is absent. Consider for the moment a generic case with $t_2 \ne t'_2$. When $t_2=0$, the sublattice symmetry is present and the dispersion curves are symmetric about zero energy, requiring the middle band of the three bands to lie at zero energy. This results in a flat band, regardless of the values of $t'_2$. The sublattice symmetry approximately holds as long as $t_2$ remains sufficiently small, which accounts for the nearly flat band observed in Fig. \ref{fig3} (a).
    

\subsection{Single trimer in presence of finite $U$}
We now analyze the gs of an isolated trimer and its wave function for the model Hamiltonian in Eq. (\ref{eq2x}) in the presence of finite $U$ at half filling. In this subsection also, we drop the index of the unit cell and simply write $c^\dagger_{i,\alpha,\sigma}$ as $c^\dagger_{\alpha,\sigma}$. The gs spin of a trimer is always $1/2$, and it is doubly degenerate, each of which is a linear combination of eight many-body configurations in the valence bond basis \cite{ramasesha1984correlated,soos1984valence}, and it can be written as 
$\ket{\Psi^\sigma_{\rm GS}}=\ket{\Psi^\sigma}+P \ket{\Psi^\sigma}$ ($\sigma=\uparrow$ or $\downarrow$) with
\begin{align}
    \ket{\Psi^\sigma} =& C_1 \ket{0}_1 \ket{\sigma}_2 \ket{\up\dn}_3 
               + C_2 \ket{\sigma}_1 \ket{\up\dn}_2 \ket{0}_3 \nonumber\\
               &+ C_3 \ket{\sigma}_1 \ket{0}_2 \ket{\up\dn}_3 
               + C_4 \ket{\sigma}_1 \ket{-}_{2,3}
\end{align}
where $\ket{-}_{2,3} = \frac{1}{\sqrt 2} (\ket{\up}_2 \ket{\dn}_3-\ket{\dn}_2\ket{\up}_3)$ and $P$ is the reflection operator with mirror plane passing through the middle site. Here $\ket{0}_\alpha$ denotes the empty state at site $\alpha$ and the other local states are defined by $\ket{\sigma}_\alpha=c^\dagger_{\alpha,\sigma}\ket{0}_\alpha$ ($\sigma=\up, \dn$) and $\ket{\up\dn}_\alpha=c^\dagger_{\alpha,\up}c^\dagger_{\alpha,\dn}\ket{0}_\alpha$.

The coefficient $C_i$ depends on $t_2$ and $U$. The magnitudes of $C_1$, $C_2$, and $C_3$ are nearly comparable in the low $U$ limit, i.e., double occupancy does not cost much. However, in the large $U$ limit, double occupancy is forbidden, and the configuration $\ket{\sigma}_1 \ket{-}_{2,3}$ and its mirror image dominate the ground state.

\subsection{Different phases}
\begin{figure}[h!]
\centering
\includegraphics[width=1.0\linewidth]{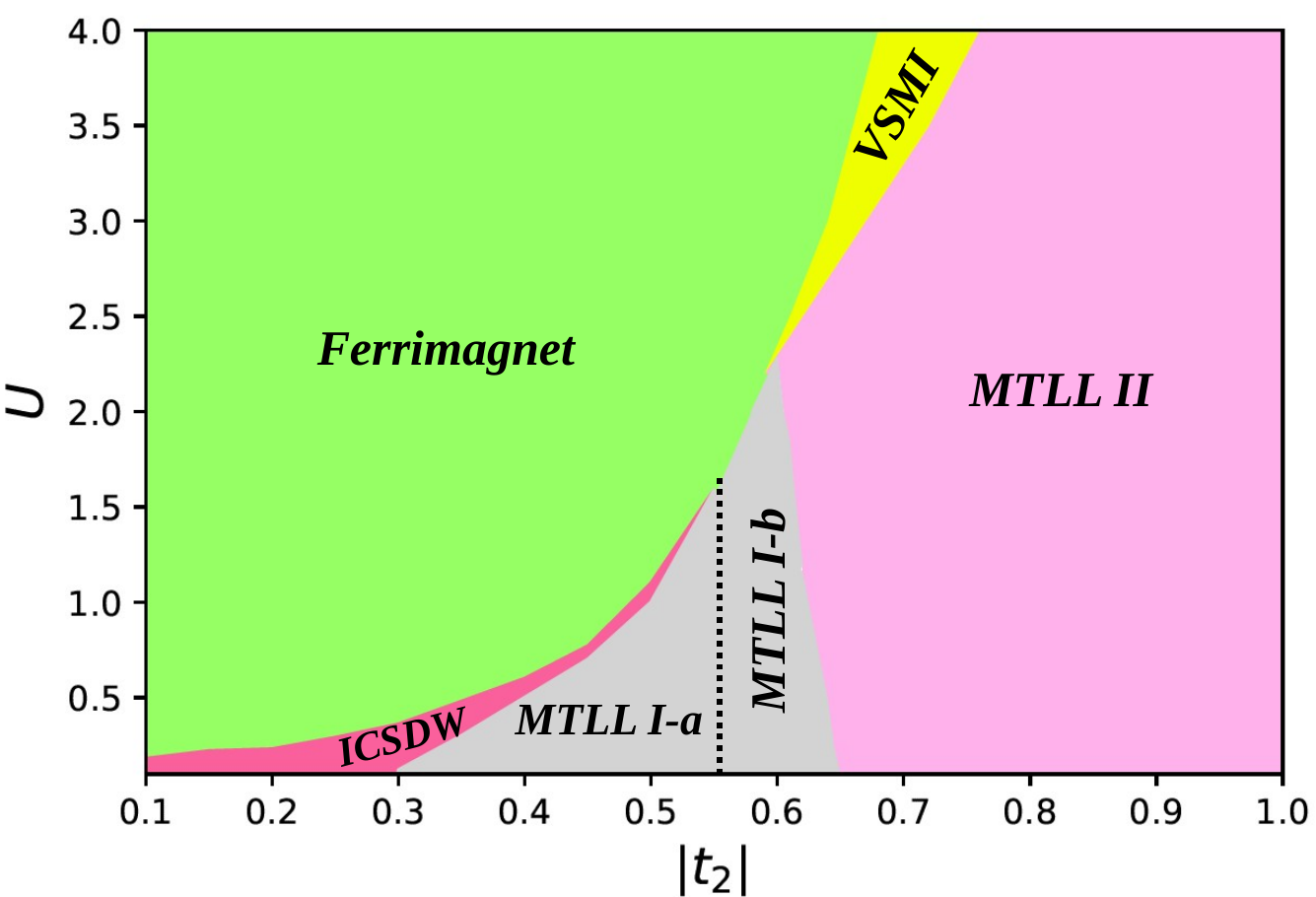}
\caption{Phase diagram of the connected trimer system using the Hubbard model at half filling. There are five phases: ferrimagnet
, insulating cell spin density wave (ICSDW), metallic Tomonaga-Luttinger liquid I (MTLL I), metallic Tomonaga-Luttinger liquid II (MTLL II) and variable spin magnetic insulator(VSMI). MTLL I is again subdivided into two parts: MTLL I-a and MTLL I-b.
 }
\label{fig5}
\end{figure}

 Competing parameters in the model Hamiltonian in Eq. (\ref{eq1}) on the geometry of connected trimers in Fig. \ref{fig1} lead to quantum fluctuations, which result in various types of quantum phases. We notice that tuning $t_2$ and $U$ at half-filling induces five quantum phases: Ferrimagnet,
  Insulating cell spin density wave (ICSDW), Metallic Tomonaga-Luttinger liquid I (MTLL I), Metallic Tomonaga-Luttinger liquid II (MTLL II) and Variable spin magnetic insulator (VSMI) phases (see Fig. \ref{fig5}). In this section, we first explain all the five phases and discuss the phase boundaries at the end of this section.  

\subsubsection{\textbf{Ferrimagnet }}

\begin{figure}[h!]
\centering{\includegraphics[width=8.8cm,height=20cm,keepaspectratio]{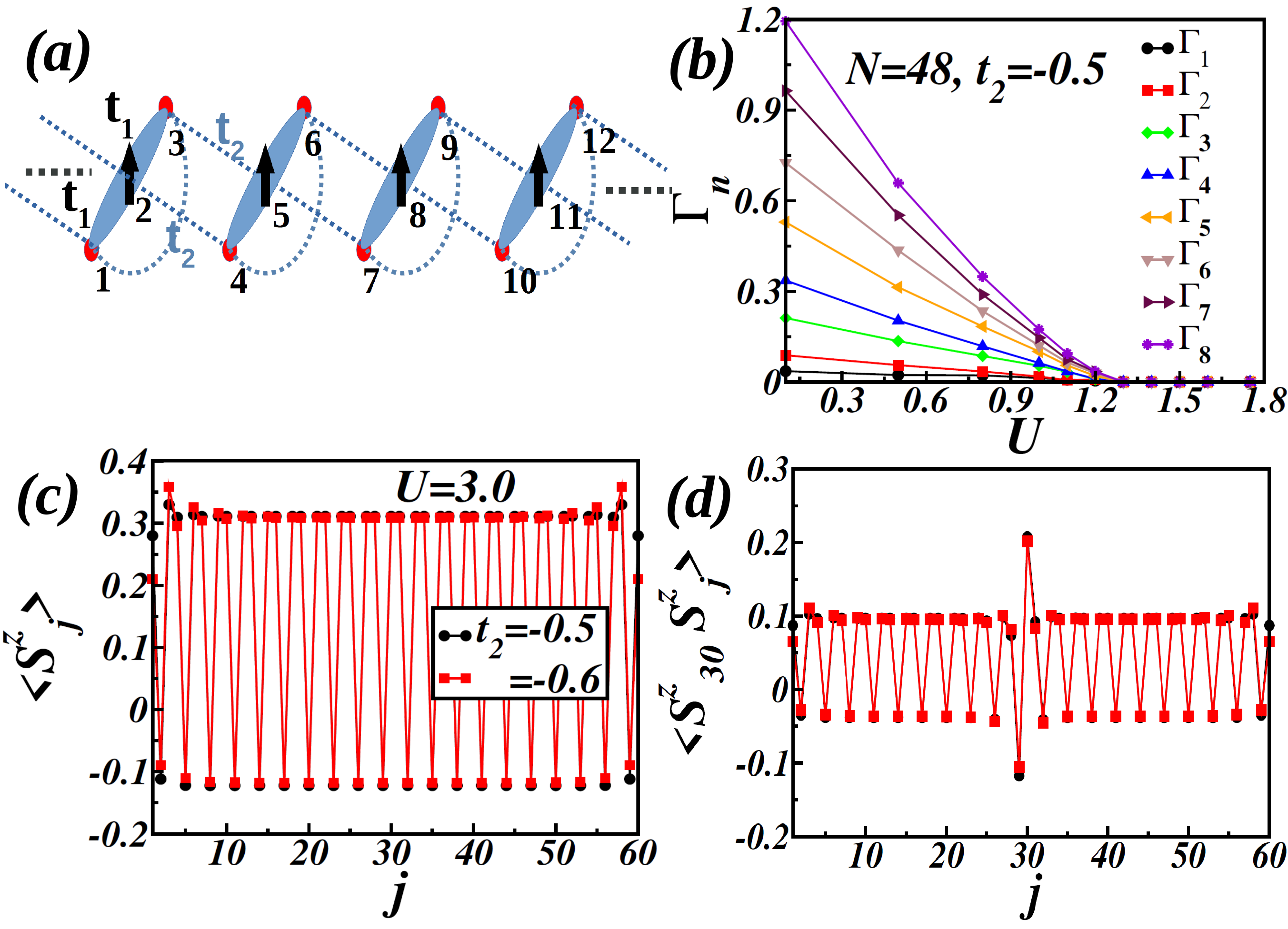}}
\caption{(a) Schematic representation of the ferrimagnetic state, where each trimer can have effective spin $1/2$ and aligns ferromagnetically. (b) Excitation gaps for a fixed $t_2=-0.5$ and varying $U$ from $0.1$ to $1.8$ for $48$ system size. (c) and (d) show the spin density and spin-spin correlation in the ferrimagnetic state for $U=3.0$ with $t_2=-0.5$ and $-0.6$, respectively. Here, $j$ is the site number from the left edge 
and is related to $(i,\alpha)$ via $j=3(i-1)+\alpha$.
 }
\label{fig6}
\end{figure}

This phase represents an insulating magnetic state where one-third of the total spins are aligned ferromagnetically, i.e., the total spin of the gs, $S_{\rm gs}$ of the system is $\frac{N}{6}$, where $N$ is the total number of sites in the system. Due to ${\rm SU}(2)$ symmetry, the ground state is $\frac{N}{3}+1$-fold degenerate.
In order to calculate the gs spin $S_{\rm gs}$, gs energies are calculated for different $z$-component of the spin, $S^z$ sectors. If the gs is in the $S_{\rm gs}$ sector, then all the lowest-energy states with $S^z \leq S_{\rm gs}$ should have the same energy.  The lowest energy gap of $S^z=n$ sector from the gs of $S^z=0$ sector 
is defined as 
\begin{eqnarray}
\Gamma_n = E_0(N, S^z=n) - E_0(N, S^z=0),
\label{eq10}
\end{eqnarray}
where $E_0(N, S^z)$ is the ground-state energy of the system in the $S^z$ sector for the number of electron $N_{\text{e}}=N$. In Fig. \ref{fig6}(b), we show $\Gamma_n$ as a function of $U$. For the given parameters $t_2 = -0.5$ and $U = 0.1$ to $1.8$, $\Gamma_n$ is finite up to $U = 1.3$ for all $n>0$. After increasing $U$ further, $\Gamma_n$ goes to $0$ for $n \le n_{\rm gs} = N/6$ and becomes finite for $n > n_{\rm gs}$. Thus, a finite $U$ is required to spontaneously break the ${\rm SU}(2)$ symmetry in the system. This critical value of $U$ depends on $t_2$; as $|t_2|$ increases, more $U$ is needed to break the symmetry due to the dispersive nature of the middle band for increasing $|t_2|$, as shown in Fig. \ref{fig5} in the phase diagram.

We can argue for the ferrimagnetic nature of the ground state \textcolor{black}{in the small $U$ and $t_2$ limit from the non-interacting band picture shown in Fig.~\ref{fig3}.} Suppose the number of electrons in the system is $N_{\text{e}}' = 2 \times (N/3)$, which is twice the number of unit cells. In this case, each $k$-point in the lowest band is filled with two electrons in the weakly interacting limit. Hence, at this filling, all possible $k$-points in the Brillouin zone are filled with two electrons with opposite spin, and thus the ground state for this filling has $S^z = 0$. However, in our case, the system is half-filled, so we must place 
remaining $(N - N_{\text{e}}')$ electrons in the higher-energy bands. These remaining $N /3$ electrons will occupy the middle, nearly flat band, which has $N/3$ possible $k$ points. 
\textcolor{black}{
Here, the situation is similar to that discussed in \cite{derzhko2014dispersion,muller2016hubbard}; the localized eigenstates do not overlap when the middle band is perfectly flat, which occurs at $t_2=0$. In this limit, the ground state remains paramagnetic and is thus highly degenerate. However, we expect that the introduction of small $t_2$ and $U$ lifts this degeneracy and causes the spins in the middle band to align ferromagnetically. This order-from-disorder mechanism has been established in a class of models discussed in \cite{derzhko2014dispersion,muller2016hubbard}. This picture naturally explains the total spin of $(N - N_{\text{e}}')/2 = N/6$ observed in the flat-band regime of the ferrimagnetic phase.} 
%
%

In large $U$ limit, this ferrimagnetic state can be explained using perturbation theory (for more details see Sec. V of the Supplemental Material \cite{suppmat}). In this phase, each trimer behaves as an effective spin-$1/2$ and ferromagnetic exchange develops between effective spin-$1/2$'s on neighboring trimers.

The insulating nature of the gs is due to the localized eigenstates of the flat band at $E_{\rm F}$. Finite magnetization in this system can be understood in terms of the Stoner criterion which suggests that $U D(E_{\rm F}) > 1$, where $D(E_{\rm F})$  is the density of states at Fermi energy of the bands \cite{stoner1938collective}. In Fig. \ref{fig6}(c) for the parameter $U=3.0$ and two different values of $t_2=-0.5$ and $-0.6$, the spin density, $\langle S^z \rangle$ in the system shows that the spins at the edge of a trimer align in the up direction, whereas the mid spin is directed down when calculated in the $S^z=N/6$ sector. Figure \ref{fig6}(d) shows the spin correlations in the $S^z = S_{\rm gs}$ sector, which exhibit a long-range order in the system, with an average magnetization $\langle S^z \rangle \approx 0.3$ at the edge of a trimer and $-0.1$ at the mid-site of it, as shown in Fig. \ref{fig6}(c).  \\

To check the metallic or insulating behavior, we calculated the charge gap, $\Delta_{\rm c}$, which is defined as \cite{ejima2007phase}
\begin{eqnarray}
\Delta_{\rm c}=\frac{1}{2}[E_0(N+2,0)+E_0(N-2,0)-2 E_0(N,0)],\hspace{0.5cm}
\end{eqnarray}
where $E_{0}(N_{\rm e},0)$ is the ground state energy of the system for the number of electrons equal to $N_{\rm e}$ in $S^z=0$ sector. 
To extrapolate $\Delta_{\rm c}$ in the thermodynamic limit, we have tested polynomials of various orders. We then found that the second-order polynomial provided the best fit to the data, which we used for the extrapolation in Fig. \ref{fig8} and \ref{fig10}.
We notice that for this phase, $\Delta_{\rm c}$ remains finite in the thermodynamic limit, as shown in Fig. \ref{fig8}(c) for the parameter $U=0.7$, $t_2=-0.4$ and Fig. \ref{fig8}(d) for $U=2.4$ and $2.5$, $t_2=-0.6$. 
\begin{figure}[b]
\centering
\includegraphics[width=1.0\linewidth]{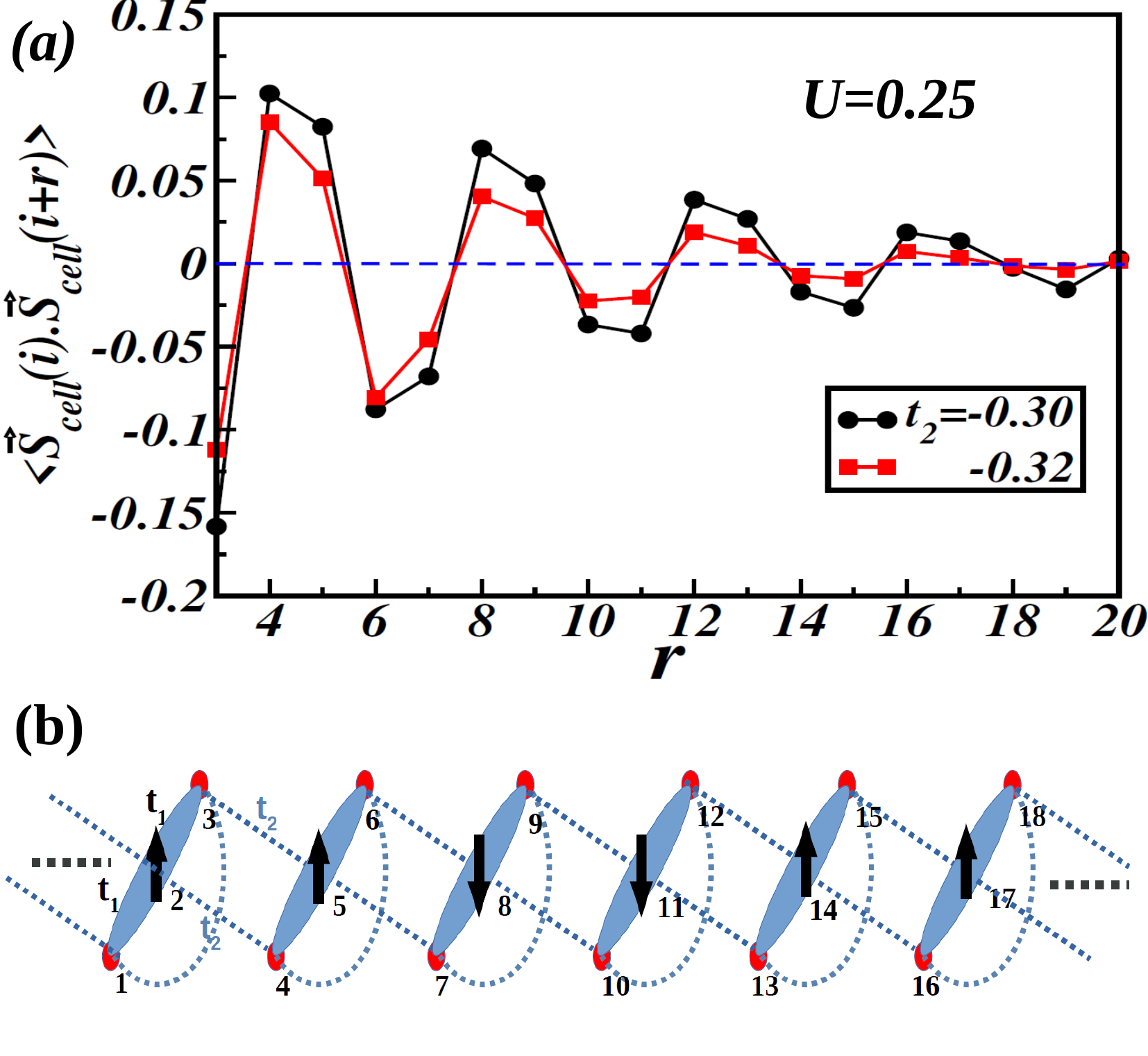}
\caption{(a) The real space spin-spin correlation between different cells for $N=120$, $U=0.25$ and $t_2=-0.3$ and $-0.32$ in the ICSDW phase. (b) Schematic representation of the ICSDW phase, where each trimer can have effective spin $1/2$ and they 
align in up-up-down-down manner. 
 }
\label{fig7}
\end{figure}

\subsubsection{\textbf{Insulating cell spin density wave (ICSDW)} }   
\begin{figure*}[hbt!]
\centering
\includegraphics[width=0.9\linewidth]{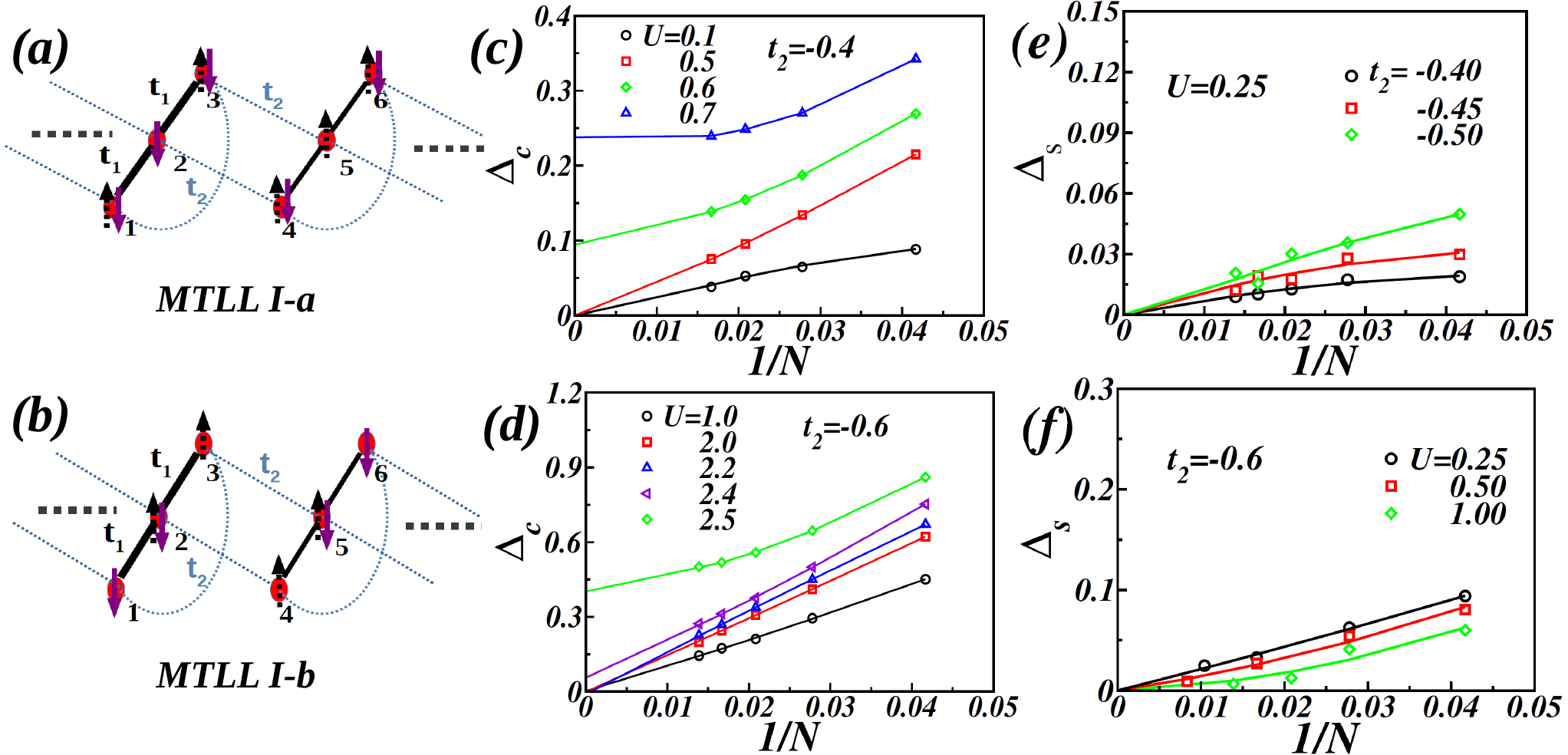}
\caption{(a) and (b) are the schematic representations of charge distribution in MTLL I-a and MTLL I-b, respectively. Here dotted spins represent the itinerant electrons in the system. Charge gaps in the thermodynamic limit are shown in (c) and (d). In Fig. (c),  $t_2$=-$0.4$ and $U=0.1$ and $0.5$ represent the MTLL I-a phase, whereas $U=0.6$ and $0.7$ correspond to the ICSDW and Ferrimagnetic
phases, respectively. In Fig. (d) $t_2=-0.6$ and $U=1.0$, $2.0$ and 2.2 corresponds to the MTLL I-b phase and $U=$2.4 and 2.5 is for the Ferrimagnetic
phase. (e) and (f) are the spin gaps in MTLL I-a and MTLL I-b, respectively.}
 
\label{fig8}
\end{figure*}

This is an insulating and nonmagnetic phase and has short-range spin and charge correlations. In this phase, the charge gap $\Delta_{\rm c}$ is finite, and this phase emerges in the regime of low $U$ and small $|t_2|$ parameters in the phase diagram. The effective spin on each trimer unit is still $S=1/2$, but it is now distributed all over the unit cell. Thus, in this case, we can define the cell spin of the $i${th} trimer as ${\bm S}_{\rm cell}(i)={\bm S}_{i,1}+{\bm S}_{i,2}+{\bm S}_{i,3}$ \cite{montenegro2006doped} . We then calculate the spin-spin correlation between different cells using the expression $\langle \bm S^{}_{\text{cell}}(i) \bm S^{}_{\text{cell}}(i+r) \rangle$. The cell spin correlation shown in Fig.\ref{fig7} (a) shows an up-up-down-down configuration.
The pictorial representation of this phase is shown in Fig. \ref{fig7} (b).  The charge gap $\Delta_{\rm c}$ for this phase is shown in Fig. \ref{fig8}(c) for the parameter $U=0.6$ and $t_2=-0.4$. Clearly, $\Delta_{\rm c}$ does not vanish in the thermodynamic limit, suggesting the insulating behavior. The insulating behavior of this phase can still be associa- 

ted with the localization of the charge due to the flat band at $E_{\rm F}$.
\subsubsection{\textbf{Metallic Tomonaga-Luttinger liquid I (MTLL I)}  }

A significant portion of the parameter space in this model contains metallic Tomonaga-Luttinger liquid (MTLL) phases. These phases are characterized as non-magnetic metals with an uneven charge distribution within the unit cell, arising from the non-bipartite nature of the geometry. Based on the charge distribution, spin excitations, and spin arrangements in the ground state, we can identify two types of MTLL phases. The first type, MTLL I, features a gapless spin and charge spectrum, quasi-long-range spin order, and an uneven charge distribution.

According to Tomonaga-Luttinger liquid (TLL) theory, the TLL parameter, $K_\rho$ is determined by the decay of the correlation functions. The charge-charge correlation function in the TLL phase  
is given by\cite{schulz1990correlation,ejima2006tomonaga} :
\begin{eqnarray}
    C^{NN}(r)\sim-\frac{K_\rho}{(\pi r)^2}+\frac{A \cos(2k_Fr)}{r^{1+K_\rho}} \ln^{-{3}/{2}}(r)  + \dots,\hspace{0.5cm}
\end{eqnarray}
where $A$ is a constant and  $K_\rho$ is the TLL parameter. The value of $K_{\rho}$ is $1$ for non-interacting fermions, whereas it is nearly $1/2$ for non-interacting spinless fermions.

The gs of the system behaves like a 1D metallic TLL and the value of $K_{\rho}$ is ranges from $1/2$ to $1$. The charge distribution within a unit cell in MTLL I can be segregated into two regions: MTLL I-a and MTLL I-b, by a cross-over. The schematic spin and charge distribution diagrams of these regions are shown in Fig. \ref{fig8}(a) and (b). The double occupancy of electrons at the edge sites and single occupancy at the mid sites characterize MTLL I-a, as shown in Fig. \ref{fig8}(a). In contrast, MTLL I-b is characterized by high electron density at the mid sites, as illustrated in Fig. \ref{fig8}(b). In this phase, electrons are delocalized through the $t_2$ hopping, which causes the flat band to become dispersive, leading to a metallic gs. 

The MTLL I phase emerges in $|t_2| >0.3$ and low $U$ regime as shown in Fig. \ref{fig5}. In this phase, the charge gap $\Delta_{\rm c}$ vanishes in the thermodynamic limit. In Fig. \ref{fig8}(c), we have shown $\Delta_{\rm c}$ as a function of $1/N$ for different values of $U$, $U = 0.1$, $0.5$, $0.6$, and $0.7$, with $t_2 = -0.4$. For $U = 0.1$ and $0.5$, the charge gap vanishes in the thermodynamic limit, corresponding to MTLL I-a. In Fig. \ref{fig8}(d),  extrapolated value of $\Delta_{\rm c}$ in MTLL I-b vanishes  for $U = 1.0$, $2.0$, $2.2$ and  $t_2=-0.6$. We have also calculated the spin gap, $\Delta_{\rm s}$ defined as:
\begin{eqnarray}
    \Delta_{\rm s}=E_0(N,S^z=1)-E_0(N,S^z=0),
\end{eqnarray}
where $E_0 (N_{\rm e}, S^z)$ denotes the gs energy of the system with $N_{\rm e}$ electrons and a total spin $S^z$ in the $z$-direction. In both the regions, $\Delta_{\rm s}$ vanishes in the thermodynamic limit. The variation of $\Delta_{\rm s}$ with $\frac{1}{N}$ in MTLL I-a for the parameter $U=0.25$ and $t_2=-0.40,-0.45$ and $-0.50$ is shown in Fig. \ref{fig8}(e).  $\Delta_{\rm s}$ in the MTLL I-b phase is shown for the parameter $t_2=-0.6$ and $U=0.25$, $0.5$, $1.0$ in Fig. \ref{fig8}(f) . 

Both the spin and the charge correlations show algebraic decay with the TLL parameter, $K_\rho$ varies from $0.5$ to $1.0$  as shown in Sec. II of the Supplemental Material \cite{suppmat}. In MTLL I-a, the charge density profile shows that it is higher at the edge sites relative to the mid site of the unit cell of a trimer, as shown in Sec. II of the Supplemental Material \cite{suppmat}. This is due to the high probability of having double occupancy at the edge sites of a trimer at low $U$ and small $|t_2|$. This region emerges at relatively low $|t_2|$ and the charge distribution over the trimer survives in finite $U$. At the higher value of $|t_2|$, $0.55 < |t_2| < 0.65$, there is a rearrangement of the charge distribution, and now the charge density is more at the mid site of the trimer (see Sec. II of the Supplemental Material \cite{suppmat}). 
 
 \begin{figure}[b]
\centering{\includegraphics[width=8.8cm,height=3.6cm]{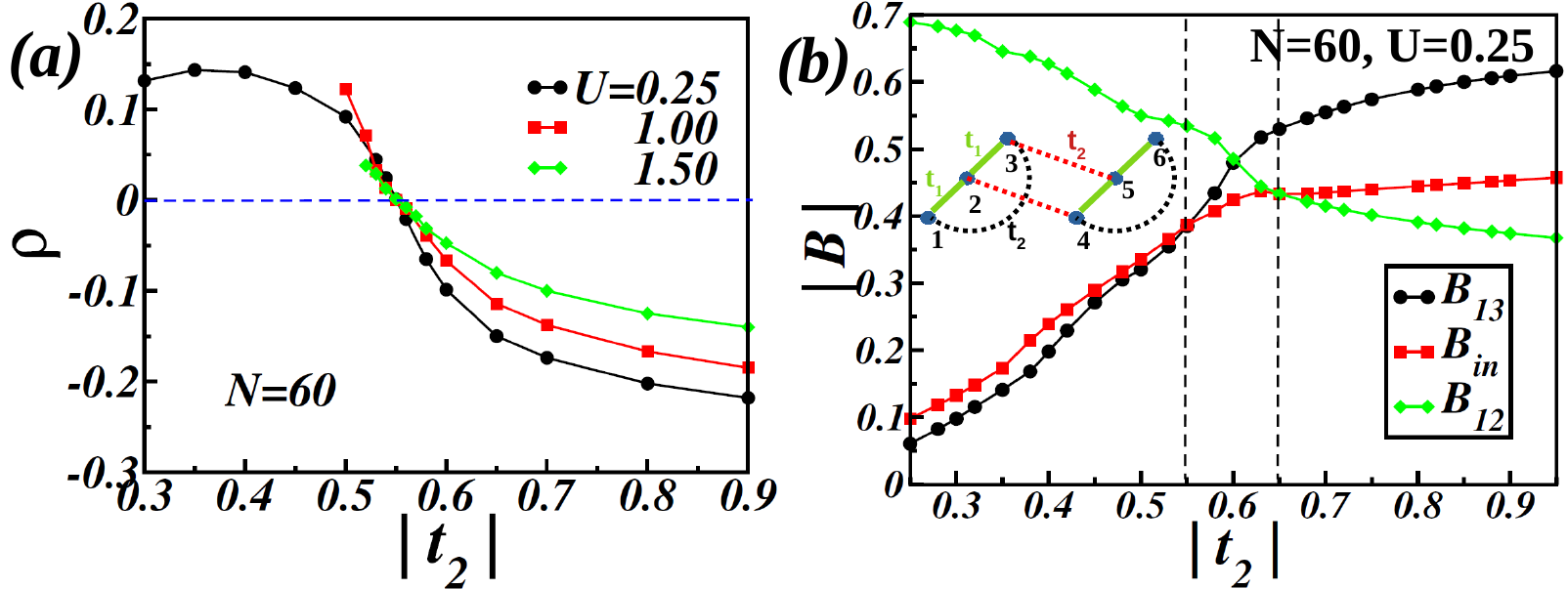}}
\caption{(a) Charge disproportionation parameter for three different values of $U=0.25$, $1.0$ and $1.5$ by varying $|t_2|$ from $0.3$ to $0.9$ for $N=60$. (b) Expectation value of the hopping term for three different type of bonds for the parameter $N=60$, $U=0.25$ and varying $|t_2|$ from $0.25$ to $0.95$. Inside this plot, two unit cells of our lattice having three different type of bonds are shown by the color black, red, and green, and corresponding to their expectation value of the hopping terms are shown in the plot by the same color.
 }
\label{fig9}
\end{figure} 

To identify the crossover between these two regions and also to show the charge rearrangement, we define a quantity called charge disproportionation $\rho$,
which can be expressed in terms of the  average charge density at the edges $\langle {\rm gs}|(n_1+n_3)|{\rm gs}\rangle$ and mid sites $\langle {\rm gs}|n_2|{\rm gs}\rangle$ in the gs as 
 \begin{eqnarray}
     \rho = \langle {\rm gs} |\frac{1}{2}(n_1 + n_3) - n_2| {\rm gs}\rangle, 
     \label{eq14}
 \end{eqnarray}
 where $|{\rm gs}\rangle$ is the ground state. 
 $\rho$ is positive in MTLL I-a and negative in MTLL I-b, as shown in Fig. \ref{fig9}(a) for three different values of $U$, $U=0.25, 1.0$, and $1.50$, by varying $t_2$. The crossover from MTLL I-a to MTLL I-b is indicated by the zero line of $\rho$. 

 Another relevant quantity to the crossover is the expectation value of the hopping term ($B$).
 Here we introduce three different types of $B$s:
   \begin{subequations}
\begin{eqnarray}
    B_{12} &=& \langle {\rm gs}|c^{\dagger}_{i,1,\sigma}c_{i,2,\sigma} + {\rm H.c.}|{\rm gs}\rangle, \label{eq15a} \\
    B_{13} &=& \langle {\rm gs}|c^{\dagger}_{i,1,\sigma}c_{i,3,\sigma} + {\rm H.c.}|{\rm gs}\rangle, \label{eq15b} \\
    B_{\rm in} &=& \langle {\rm gs}|c^{\dagger}_{i,1,\sigma}c_{i-1,2,\sigma} + {\rm H.c.}|{\rm gs}\rangle. \label{eq15c}
\end{eqnarray}
\end{subequations}
 Expectation value of the hopping term between the edge and middle site, $B_{12}$, is given by Eq. (\ref{eq15a}),  and Eq. (\ref{eq15b}) represents the expectation value of the hopping term between the two edge sites in the same $i$th trimer unit cell. The inter-trimer $B$ is denoted by $B_{\rm in}$ which is given in Eq. (\ref{eq15c}). We choose $i$ to be the unit cell at the center of the ladder and compute these three types of $B$s. The results are shown in Fig. \ref{fig9}(b) for $N=60$ and $U$=$0.25$. In the MTLL I-a, $B_{12}$ is the largest $B$, and $B_{\rm in}$ is stronger compared to $B_{\rm 13}$. There is a crossover between $B_{13}$ and $B_{\rm in}$ at $t_2 \approx -0.55$ and it is consistent with the crossover obtained from $\rho$.

\subsubsection{\textbf{Metallic Tomonaga-Luttinger liquid II (MTLL II)} 
} 
\begin{figure}[hb]
\centering{\includegraphics[width=8.8cm,height=20cm,keepaspectratio]{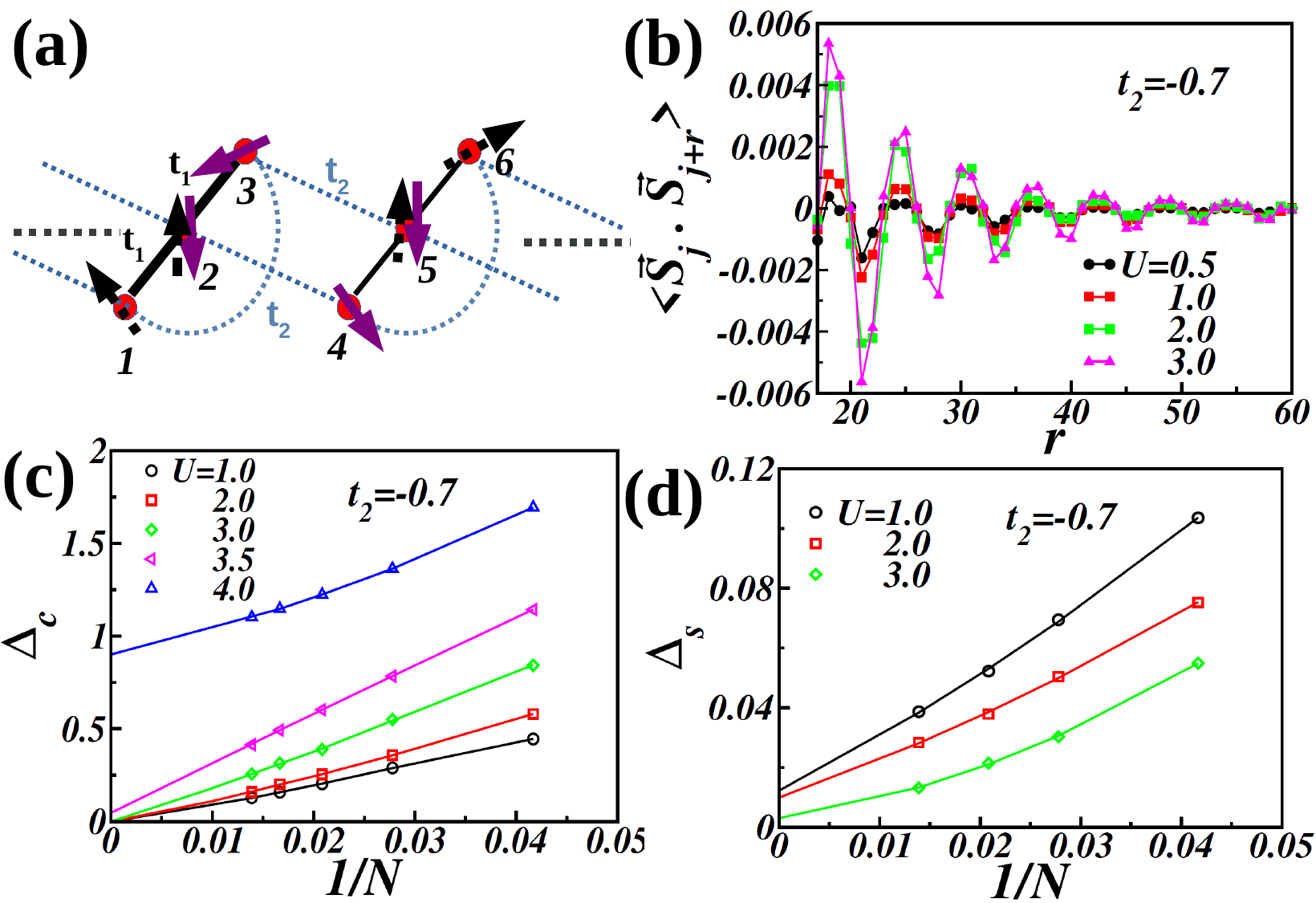}}
\caption{(a) Schematic representation of the MTLL II phase. Dotted spins represent the itinerant electrons moving in the system. (b) Spin-spin correlation in 
the MTLL II phase for the parameter $N=120$, $t_2=-0.7$ and $U=0.5$, $1.0$, $2.0$, and $3.0$. (c) and (d) show the finite-size scaling analysis of the charge gap and the spin gap in the MTLL II phase, respectively.
 }
\label{fig10}
\end{figure}

It is also a metallic phase with short-range spin order or spiral order in the gs. This phase is another type of TLL phase with a gapless charge mode but a gapped spin mode. The spin correlation shows short-range spiral behavior that arises due to spin frustration induced by the competing nearest-neighbor hopping $t_1$ and next-nearest-neighbor hopping $t_2$ in the system. The schematic representation of this phase is shown in Fig. \ref{fig10}(a). 
The pitch angle of this spin spiral is roughly constant on tuning $U$ and $t_2$ as shown in Fig. \ref{fig10} (b) and Sec. III of the Supplemental Material. In the thermodynamic limit, the charge gap  $\Delta_{\rm c}$ is zero, which indicates the metallic behavior of the gs. The extrapolation of $\Delta_{\rm c}$ is shown as a function of $1/N$ in Fig. \ref{fig10}(c) for $t_2=-0.7$ and $U=1.0$, $2.0$ and $3.0$. 
In Fig. \ref{fig10}(d), the spin gap, $\Delta_{\rm s}$, is extrapolated as a function of $1/N$ in this phase. In the thermodynamic limit, the spin gap remains finite, which may be because of underplaying frustration in the system. This phase emerges for large $|t_2|$ and survives up to the maximum value of $U=4$ studied in this work. The charge-charge correlation decays algebraically with the TLL parameter, $K_\rho$ lying between $0.5$ and $1.0$, and is more dominant compared to the exponentially decaying spiral spin-spin correlation. The charge density distribution in this phase is shown in Fig. S3(b) in the Supplemental Material. The charge density is higher at the mid sites compared to the edge sites of the trimer as shown in Fig. \ref{fig9}(a). The phase boundary between MTLL I-b and MTLL II can also be predicated by the crossover of expectation value of the hopping term $B_{12}$ and $B_{\rm in}$. The expectation value of edge sites hopping term, $B_{13}$ is the largest, and the inter-trimer hopping term, $B_{\rm in}$ is larger than $B_{12}$ in this phase as shown in Fig. \ref{fig9}(b).

\subsubsection{\textbf{Variable spin magnetic insulator (VSMI)}} 

\begin{figure}[hbt!]
\centering
\includegraphics[width=1.0\linewidth]{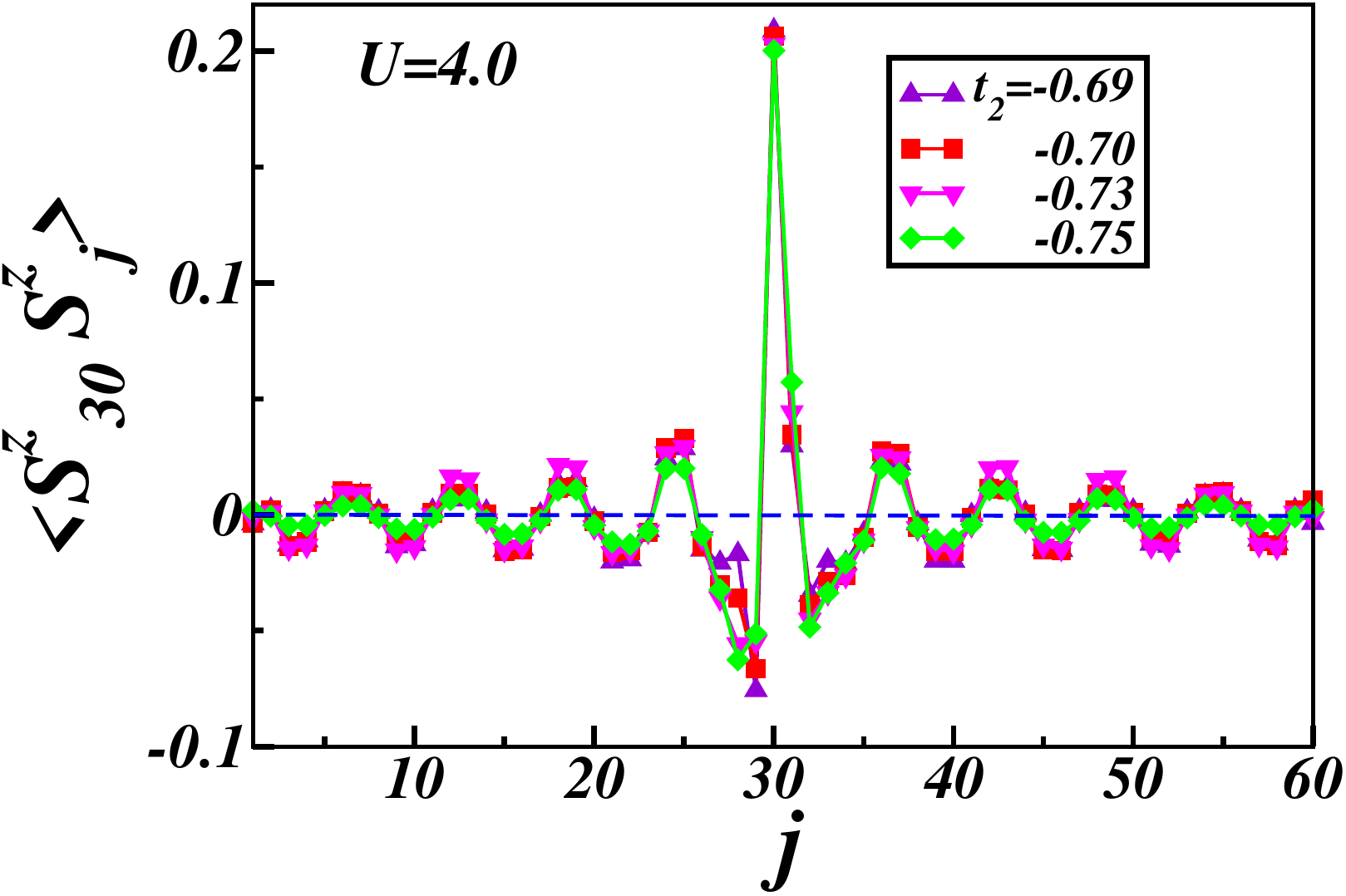}
\caption{Real space spin-spin correlation in VSMI phase for the parameter $N=60$, $U=4.0$ and $t_2=-0.69$, $-0.70$, $-0.73$, and $-0.75$ by taking the mid site as a reference.
Here, $j$ is the site number from the left edge and is related to $(i, \alpha)$ via $j = 3(i - 1) + \alpha$.}
 
\label{fig12}
\end{figure}

This phase is an insulating high spin state with the total spin of the gs, $S_{\rm gs}$ less than $N/6$ and this phase emerges in the parameter range $0.65<|t_2| <0.75$ and  $2 < U <4$.  This phase also shows quasi-long-range spiral order correlation which is shown in $S^z=0$ sector in Fig. \ref{fig12}. 

\begin{table*}[t]
    \centering
   \caption{Criterion for 
    determining phase boundaries between different phases. } 
Here, \( S_{\text{gs}} \) represents the ground state spin of the system, \( \Delta_{\rm c} \) denotes the charge gap, and \( \rho \) refers to the charge disproportionation
as defined in Eq.~(\ref{eq14}). The terms $B$ and \( \langle \bm{S}_j \cdot \bm{S}_{j+r} \rangle \) correspond to the expectation value of hopping terms and spin-spin correlation, respectively.

    \begin{tabular}{|>{\centering\arraybackslash}p{0.15\linewidth}|>{\centering\arraybackslash}p{0.15\linewidth}|>{\centering\arraybackslash}p{0.08\linewidth}|>{\centering\arraybackslash}p{0.08\linewidth}|>{\centering\arraybackslash}p{0.15\linewidth}|>{\raggedright\arraybackslash}p{0.16\linewidth}|} \hline  
         
Phase
&  $S_{\rm gs}$&  $\Delta_{\rm c}$&  $\rho$& $B$&  $\langle{\bm S}_{j}\cdot{\bm S}_{j+r}\rangle$\\ \hline  
         Ferrimagnet&  $N/6$&  $> 0$&  --& -- &  long-range\\ \hline  
         ICSDW&  $0$&  $> 0$&  --& -- & short-range\\ \hline  
         MTLL I-a&  $0$&  $0$&  $>0$& $B_{13}\approx B_{in} < B_{12}$& quasi long-range \\ \hline  
         MTLL I-b&  $0$ & $0$ &  $<0$& $B_{13}\approx B_{in}\approx B_{12}$& quasi long-range \\ \hline  
         MTLL II&  $0$&  $0$&  $<0$& $B_{12} < B_{in} < B_{13}$& short-range\\ \hline  
         VSMI&  $0<S_{\rm gs}<N/6$&  $> 0$&  --& -- & quasi long-range\\ \hline 
    \end{tabular}
    
    \label{tab:my_label}
\end{table*}
Here $S_{\rm gs}$ is sensitive to the parameters and it decreases with increasing the magnitude of $t_2$ for fixed $U$. However, it increases with increasing $U$ for a given value of $t_2$. In Fig. S4(a) and (b) in the Supplemental Material,  $\Gamma_{n}$ (defined in Eq. (\ref{eq10})) is shown for two system sizes and based on that, we determine $S_{\rm gs}$ for $N=36$ which is also shown as a function of $|t_2|$ for $U =4.0$ in the same Fig. S4 (c). $\Delta_{\rm c}$ in Fig. \ref{fig10}(c) for the parameter $t_2=-0.7$ and $U=3.5$ and $4.0$ show a finite value in the thermodynamic limit, which indicates the insulating behavior of this phase.

\subsection{Phase boundaries}
In Fig.\ref{fig5} we show the quantum phase diagram in $|t_2|$ and $U$ parameter space. The boundaries of these phases are determined by calculating various quantities listed in Table \ref{tab:my_label}. The phase boundary between ferrimagnet
and ICSDW is calculated based on the effective value of $S_{\rm gs}$, the spin-spin correlations, and the cell spin correlation in Fig. \ref{fig7}. The phase boundary between ICSDW and MTLL I-a is determined based on $\Delta_{\rm c}$. The charge disproportionation
$\rho$ and expectation value of hopping term $B$ are used to distinguish MTLL I-a from MTLL I-b. The boundary between ferrimagnet
and MTLL I-b is determined based on the $S_{\rm gs}$ and charge gap $\Delta_{\rm c}$ calculations. The boundary between MTLL I-b and MTLL II is determined based on the spin-spin correlations and the $B$ calculation. The boundary between MTLL II and VSMI is determined by $S_{\rm gs}$ and the $\Delta_{\rm c}$, whereas the boundary between ferrimagnet
and VSMI is determined using the $S_{\rm gs}$ calculation and the spin-spin correlation.

\section{DISCUSSION AND SUMMARY}

In this manuscript, we have studied the Hubbard model on a coupled trimer geometry, as illustrated in Fig. \ref{fig1}. We examined how the  competing hopping parameters $t_1$, $t_2$, and the Hubbard interaction $U$ affect the ground state properties of the system. We used the exact diagonalization (ED), density matrix renormalization group (DMRG), and perturbation theory to construct a quantum phase diagram in the $t_2$-$U$ plane, 
where $t_1$ sets the energy scale in the system. Our findings reveal five distinct quantum phases within this parameter space. At low $U$ and moderate to high values of $t_2$, the system exhibits metallic behavior, while for large $U$, the ground state becomes a magnetic insulator.

In the moderate to high $U$ and up to a certain magnitude of $t_2$, the ground state is insulating and magnetic. In this regime, the effective spin of the ground state is one-third of the total spin. In small $U$ and small $t_2$ limit, this magnetic gs can be explained from the flat band ferromagnetism as in our case the nearly flat band is situated at the middle of the dispersion curve. However, in the large $U$ limit, a ferrimagnetic state can arise from the formation of resonant singlet dimers between nearest-neighbor spins and free spins on each trimer. From the perturbation theory, we see that these free spins are coupled via ferromagnetic exchange with their nearest neighbors as shown in Fig. \ref{fig6}(a). This phase closely resembles that studied by Giri et al. \cite{giri2017quantum}, where the spins on each trimer behave as an effective spin $S=1/2$ and are coupled ferromagnetically to neighboring effective spins. Such a phase can also be observed in insulating and distorted azurite systems like Cu$_3$(P$_2$O$_6$OH$)_2$ \cite{shao2023progress,montenegro2022ground}, and in the one-third magnetic plateau phases in weakly coupled trimer systems in Na$_2$Cu$_3$Ge$_4$O$_{12}$ under a finite magnetic field \cite{bera2022emergent,yasui2014magnetic}.

At small $U$ and small magnitude of $t_2$, the spin-1/2 magnetic moments become delocalized over the trimer. The competition between $t_1$ and $t_2$ induces the ICSDW phase, characterized by an up-up-down-down spin configuration. In the moderate to large magnitude of $t_2$ regime, the ground state of the system behaves as metal, and it exhibits Tomonaga-Luttinger liquid behavior. Depending on the lowest spin gap, we classify this phase into two types: MTLL I, which has gapless spin excitations, and MTLL II, which features a gapped spin excitations and a spiral spin structure in the ground state. The uneven charge distribution between edge and middle sites further divides MTLL I into MTLL I-a and MTLL I-b. As the magnitude of $t_2$ increases, the system remains metallic. In the MTLL II phase, the competition between hopping amplitudes at finite $U$ leads to spin frustration, resulting in a spin gap. At high $U$, the ground state goes from a ferrimagnetic
state to a variable spin magnetic insulator (VSMI) state as the magnitude of $t_2$ increases, where the frustration due to increasing $t_2$ causes the melting of the ferrimagnetic ground state and a reduction in ground-state spin.

In summary, we have investigated a Hubbard model on a trimer ladder geometry and demonstrated that by tuning competing hopping and onsite Coulomb interaction $U$, five distinct quantum phases arise in the gs of the system. While trimer ladder systems are naturally abundant, they are mostly insulating. However, some of these materials under pressure, could exhibit variable spin behavior at low temperatures, as well as intriguing metallic properties characteristic of Tomonaga-Luttinger liquid. We hope that this work will attract the interest of experts in the field and prompt further investigation of these materials.

\section*{ACKNOWLEDGMENT}
S.S. thanks DST INSPIRE for the financial support. M.K. thanks DST for funding through grant no. CRG/2020/000754. H.K. was supported by JSPS KAKENHI Grants No. JP23K25783, No. JP23K25790, and MEXT KAKENHI Grant-in-Aid for Transformative Research Areas A “Extreme Universe” (KAKENHI Grant No. JP21H05191).
\par

\medskip

\bibliography{reference}

\end{document}


\title{Phase diagram of a coupled trimer system at half filling using the Hubbard model } 
\author{Sourabh Saha}
\affiliation{Department of Condensed Matter and Materials Physics,
S. N. Bose National Centre for Basic Sciences, JD Block, Sector III, Salt Lake, Kolkata 700106, India}
\author{Hosho Katsura}
\email{katsura@phys.s.u-tokyo.ac.jp}
\affiliation{Department of Physics, The University of Tokyo, 
Hongo, Bunkyo-Ku, Tokyo 113-0033, Japan} 
\affiliation{Institute for Physics of Intelligence, The University of Tokyo, 
Hongo, Bunkyo-ku, Tokyo 113-0033, Japan} 
\affiliation{Trans-scale Quantum Science Institute, The University of Tokyo, Hongo, Bunkyo-ku, Tokyo 113-0033, Japan}
\author{Manoranjan Kumar}
\email{manoranjan.kumar@bose.res.in}
\thanks{Last two authors contributed equally.}
\affiliation{Department of Condensed Matter and Materials Physics,
S. N. Bose National Centre for Basic Sciences, JD Block, Sector III, Salt Lake, Kolkata 700106, India}

\begin{abstract}
	Here we provide supplemental explanations and data on the following topics in relation to the main text: I. Mapping of trimer ladder to other geometries.
 II. Charge density and charge-charge correlation in the MTLL I phase.
 III. Contains the information about spin-spin, charge-charge correlation and charge density in the MTLL II phase. In IV., we discuss the GS spin in the VSMI phase. V. Contains the details of perturbation theory.
\end{abstract}
\date{\today}
\maketitle

\section{\label{sec1} Mapping of trimer ladder to other geometries } 
\begin{figure}[h!]
\renewcommand{\thefigure}{S1}
    \centering
    \includegraphics[width=0.6\linewidth]{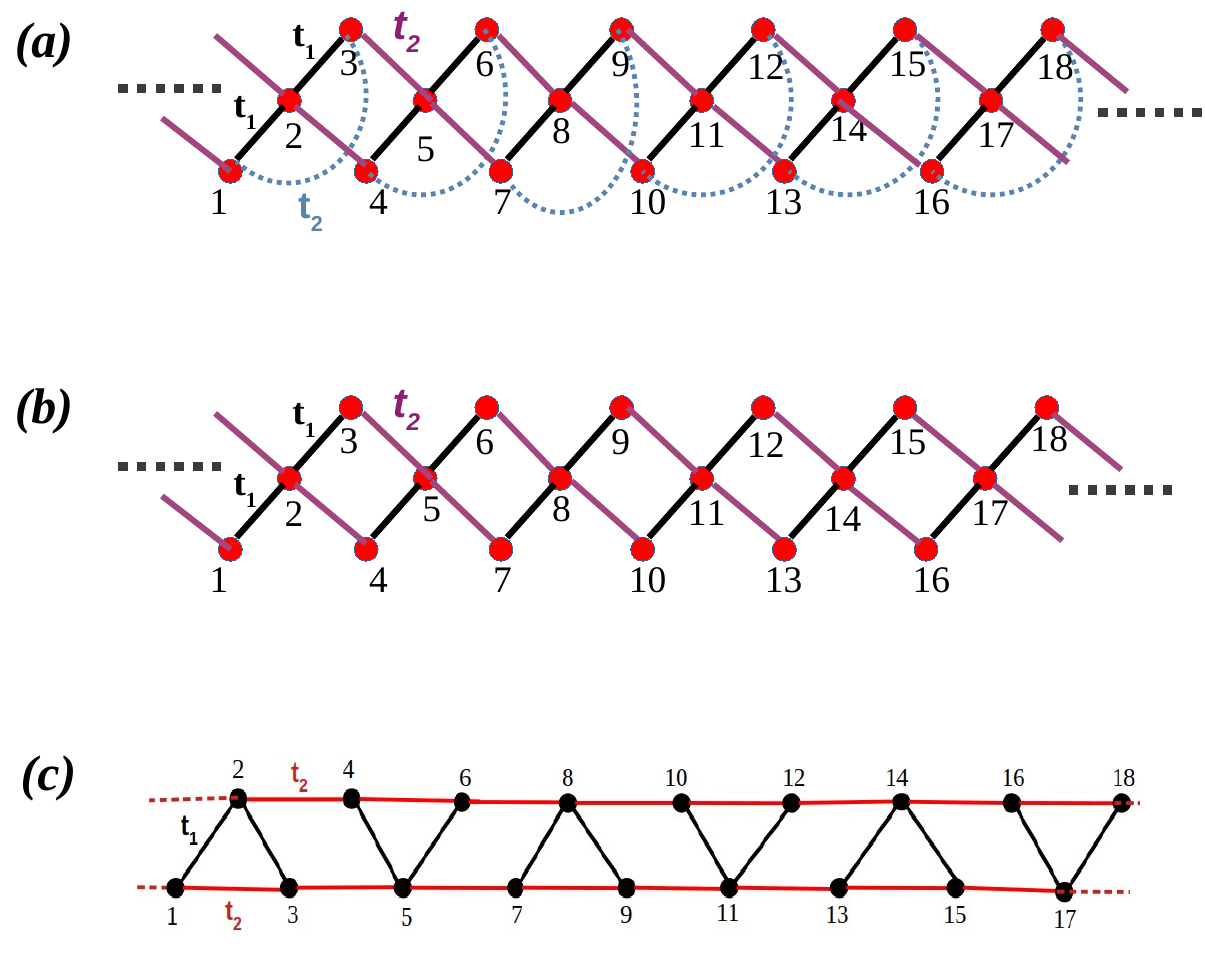}
    \caption{(a) Schematic representation of 
    coupled trimer system is shown with the hopping parameters $t_1$ and $t_2$. (b)  On removal of the next-nearest-neighbor $t_2$ within each trimer in (a), the structure resembles a diamond lattice. (c) Bending the trimer into a triangle and connecting these trimers leads to $3/4$ structure as in Fig. 1(b) of Ref. [\onlinecite{giri2017quantum}].} 
    \label{S1}
\end{figure}

In this section, we show that coupled trimer geometry as shown in Fig. \ref{S1}(a) can be mapped to various geometries like distorted diamond lattice \cite{shahbazi2023revival} and  $3/4$ ladder geometries \cite{giri2017quantum}. In Fig. \ref{S1}, we show that removing the next nearest neighbor in each trimer leads to a diamond lattice as shown in Fig. \ref{S1}(b). If we fold 
two nearby trimers into two triangles and connect it with inter-trimer bonds, then the structure looks like a $3/4$ ladder geometry as shown in Fig. \ref{S1}(c). 

{\label{sec2}

\section{Charge Density and Charge-Charge Correlation in MTLL I phase}
In this section, we present the real space charge density and charge-charge correlations in two regions of MTLL I phase : MTLL I-(a) and MTLL I-(b).
Fig. \ref{S2}(b) depicts the charge density profile at different sites $j$ for MTLL I-(a), for the parameters $U=0.25$, and $t_2 = -0.40$, $-0.45$, and $-0.50$ and $N=120$ system size. The results show that the charge density exceeds 1.0 at the corner sites of the trimer, while it is approximately 0.92 at the mid-site, indicating fluctuations in charge density in this phase. Here charge-charge correlation decays algebraically with distance $r$ which is shown in Fig. \ref{S2}(c) along the chain (as shown in Fig.\ref{S2}(a) by the green line) on a log-log scale and fitted it with the expression given in Eq. (12) in the main text for the parameters discussed earlier. For the MTLL I-(b), charge density is higher at the mid-sites of the trimer compared to the end sites, as shown in Fig. \ref{S2}(d). In this region, the charge-charge correlation also demonstrates a quasi-long-range order, and it is shown 
on a log-log scale along the chain as shown in Fig.\ref{S2}(a) and it follows the power law behaviour with LL parameter varies from $0.5$ to 1.0, as shown in Fig. \ref{S2}(e).

\begin{figure}[h!]
\renewcommand{\thefigure}{S2}
    \centering
    \includegraphics[width=0.8\linewidth]{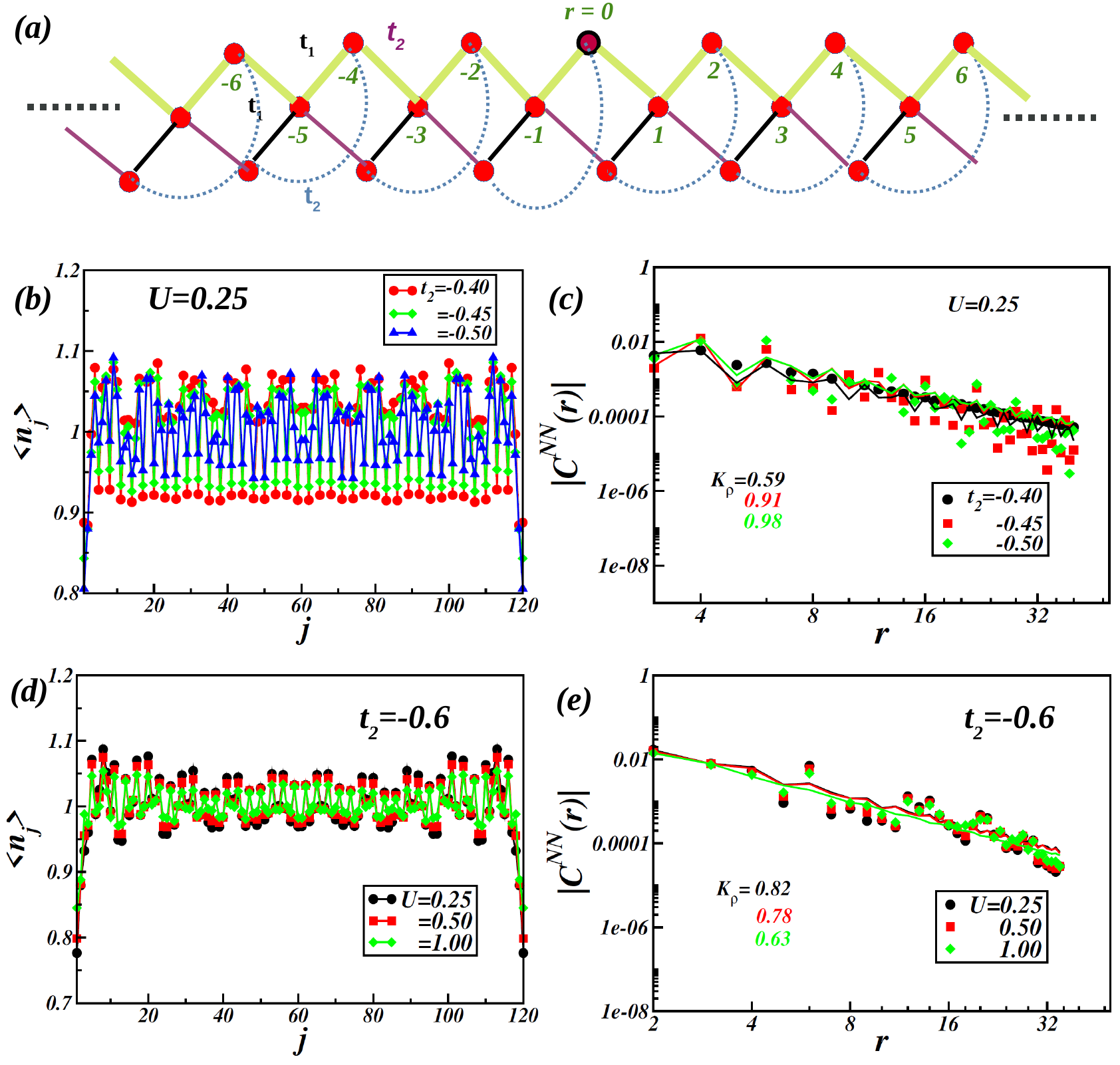}
    \caption{
 (a) show the distance taken from the reference at $r=0$ to various r by the green line for calculating the charge-charge correlation in MTLL I-a, MTLL I-b and MTLL II.   (b) and (c) show the charge density at site $j$ and charge-charge correlation along a chain at distance $r$ in  MTLL I-a for the parameter $N=120,U=0.25$ and $t_2=-0.40$, $-0.45$, and $-0.50$, respectively. Similarly, charge density and charge-charge correlation along a chain for  MTLL I-b is shown in (d) and (e), respectively, for the parameters $N=120$, $t_2=-0.6$ and $U=0.25$, $0.5$ and $1.0$.}
    \label{S2}
\end{figure}

\section{Spin-spin, charge-charge  correlation and charge density in MTLL II phase}
Spin-spin correlations in the metallic Tomonaga-Luttinger liquid II  (MTLL II) phase are shown in Fig. \ref{S3}(a) for a fixed $U$, $U=3.0$, by varying $t_2$ from $-0.7$ to $-0.85$ in units of $-0.05$. The spin-spin correlation shows spiral behavior but the pitch angle remains almost constant. Fig. \ref{S3}(b) shows the real space charge density profile with the site index $j$ for the parameter $t_2$=$-0.7$ and $U=0.5,1.0,2.0$ and $3.0$. In this phase also the charge density is more than $1.0$ at the mid site of a trimer and it is less than 1.0 in the two corner sites of it. Charge-charge correlation, $C^{NN}(r)$ along a chain, which is shown in Fig.\ref{S2}(a) by the green line, shows algebric decay with the LL parameter varies between $0.5$ to $1.0$ as shown in Fig. \ref{S3}(d) for the parameter $t_2=-0.7$ and $U=1.0, 2.0, 3.0$.
\begin{figure}[h!]
\renewcommand{\thefigure}{S3}
    \centering
    \includegraphics[width=1.0\linewidth]{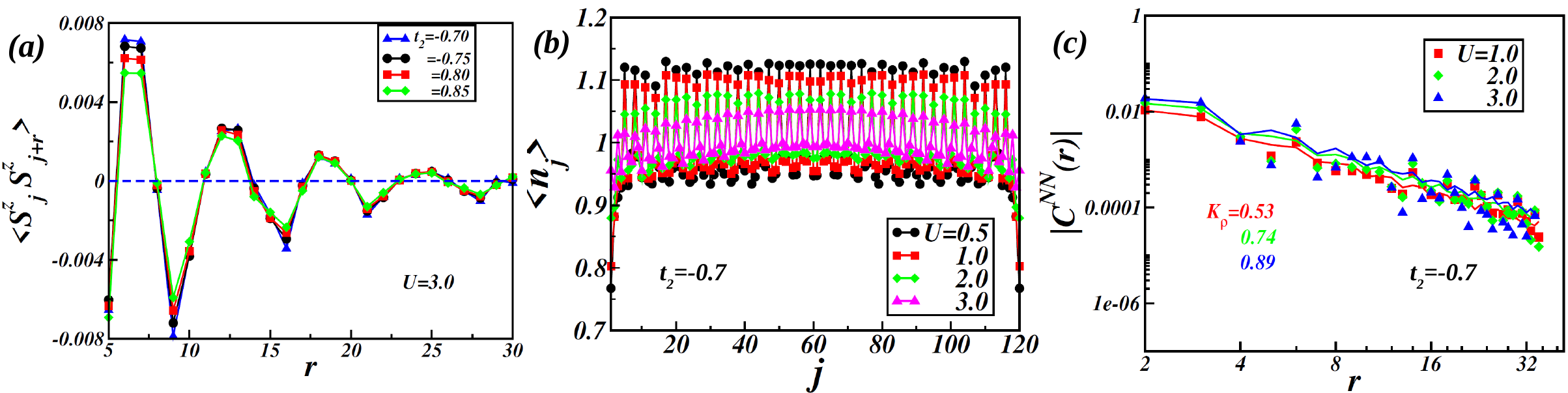}
    \caption{(a) Spin-spin correlation in the MTLL II phase for fixed value of $U$, $U=3.0$ and different values of $t_2=-0.70$, $-0.75$, $-0.80$, and $-0.85$. (b) Charge density for the parameters $t_2=-0.7$ and $U=0.5$ , $1.0$, $2.0$, and $3.0$ in the MTLL II phase. (c) Charge-charge correlation along a chain in MTLL II phase for the parameter $t_2=-0.7$ and $U=1.0,2.0,3.0$.}
    \label{S3}
\end{figure}

\section{GS spin in the VSMI phase}
In Fig. \ref{S4}(a) and (b), we 
show the variation of energy gap $\Gamma_n$ (as given in Eq. (10) in the main text) for different values of $n$ with $|t_2|$. It is seen that for a fixed $U$, $U=4.0$, as we vary $|t_2|$, the system goes from the FOTS phase, where the gs spin of the system is $S_{gs}=N/6$, to the VSMI phase, where $S_{gs}$ 
changes with $t_2$ as shown in Fig. \ref{S4}(a) for $N=24$ and \ref{S4}(b) for $N=36$. In Fig. \ref{S4}(c), the variation of $S_{gs}$ with $|t_2|$ is shown for the parameter $N=36$, $U=4.0$.
\begin{figure}[h!]
\renewcommand{\thefigure}{S4}
    \centering
    \includegraphics[width=1.0\linewidth]{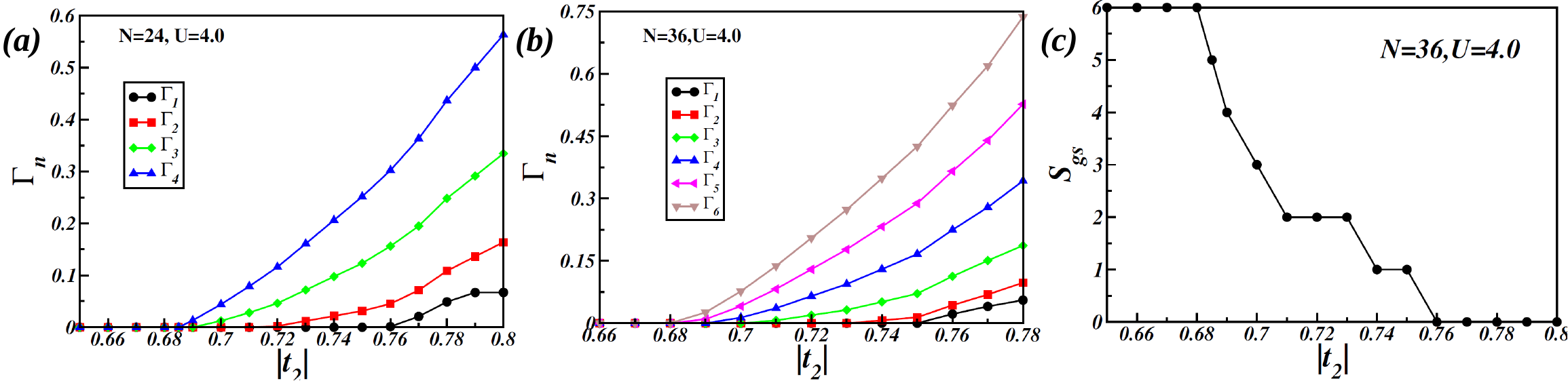}
    \caption{(a) and (b) are the variation of  energy gap, $\Gamma_n$ for different values of $n$  with $|t_2|$ for a fixed $U=4.0$ and $N=24$ and $36$ respectively, in the VSMI phase. (c) Ground state spin, $S_{gs}$ with $|t_2|$ for $N=36,U=4.0$.}
    \label{S4}
\end{figure}
\section{Perturbation theory }

\begin{figure}
    \centering
    \renewcommand{\thefigure}{S5}
    \includegraphics[width=0.8\linewidth]{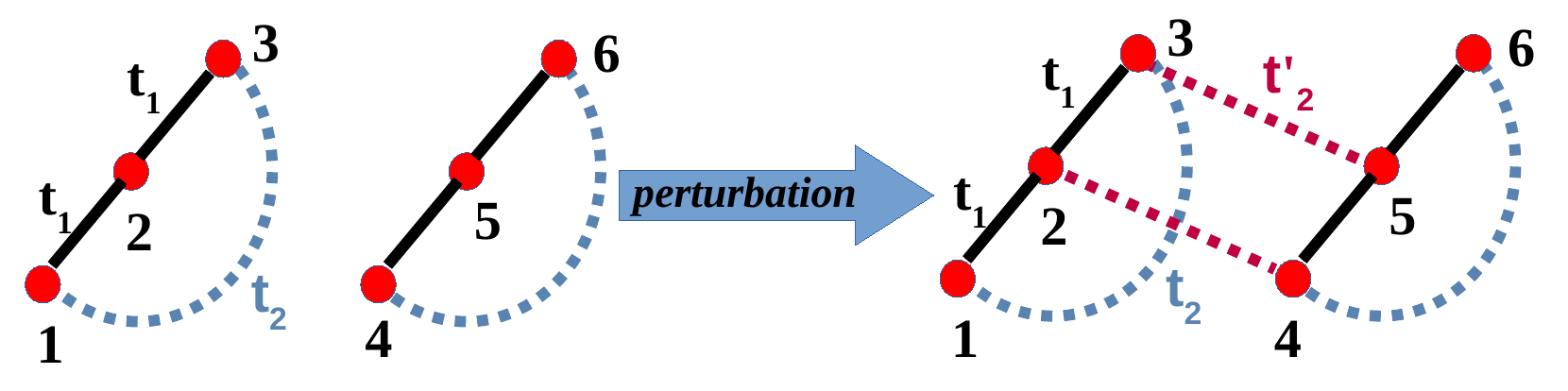}
    \caption{ On the left, two unperturbed trimer unit cells are shown. The perturbation, the inter-trimer hopping $t'_2$, is included on the right. Here, $t_1$ and $t_2$ are the amplitudes of the nearest and next-nearest neighbor hopping within the same trimer, respectively.} 
    \label{S5}
\end{figure}

In this section, we analyze the perturbation theory for our coupled trimer system by considering two trimer unit cells. 
A similar approach can be found in Refs. \onlinecite{ye2024kinetic,janani2014low}.
The unperturbed Hamiltonian consists of two decoupled trimer unit cells, with each trimer treated individually. The unperturbed Hamiltonian for a single trimer unit cell is expressed as
\begin{equation}
H_{\text{trimer}} = t_1 \sum_{\sigma=\uparrow,\downarrow} \left( c_{1,\sigma}^\dagger c_{2,\sigma} + c_{2,\sigma}^\dagger c_{3,\sigma} + \text{H.c.} \right) + t_2 \sum_{\sigma=\uparrow,\downarrow} \left( c_{1,\sigma}^\dagger c_{3,\sigma} + \text{H.c.} \right) + U \sum_{i=1}^3 n_{i,\uparrow}n_{i,\downarrow},
\end{equation}
where \(t_1\)($<0$) is the hopping parameter between nearest neighbor sites, and \(t_2\) ($<0$) is the hopping between next-nearest neighbor sites within the same trimer, as illustrated in Fig. \ref{S5}. The term \(U\)($>0$) represents the on-site Coulomb interaction between electrons of opposite spin on the same site.

At half-filling (with six sites and six electrons), the ground state configuration has three electrons per trimer rather than unevenly distributed electrons, like four electrons in one trimer and two in the other. In this configuration, each trimer behaves effectively as a spin-$1/2$ system, with a two-fold degenerate ground state, denoted by \(\ket{\Uparrow}\) and \(\ket{\Downarrow}\).

For the unperturbed case, 
the ground states of the full system consisting of two trimers are
\begin{align*}
    & \ket{\Uparrow}\ket{\Uparrow} \\
    & \ket{\Downarrow}\ket{\Downarrow} \\
    & \frac{1}{\sqrt{2}} \left( \ket{\Uparrow}\ket{\Downarrow} + \ket{\Downarrow}\ket{\Uparrow} \right) \\
    & \frac{1}{\sqrt{2}} \left( \ket{\Uparrow}\ket{\Downarrow} - \ket{\Downarrow}\ket{\Uparrow} \right).
\end{align*}
The first three states correspond to the triplet, while the last one corresponds to the singlet. Thus, the total unperturbed system at half-filling has a four-fold degenerate ground state. 
We treat the effect of inter-trimer hopping $t'_2$ in degenerate perturbation theory. 

To investigate the high-energy part of the unperturbed system, we calculate 
the eigenenergies and eigenstates of the single-trimer Hamiltonian $H_{\rm trimer}$ in different particle-number sectors: \(N_{\rm e} = 2\) and \(N_{\rm e} = 4\). 
Let $\ket{\Psi_{2,m}}$ be the $m$th (normalized) eigenstate of $H_{\rm trimer}$ in the $N_{\rm e}=2$ sector. Similarly, let $\ket{\Psi_{4,n}}$ be the $n$th (normalized) eigenstate of $H_{\rm trimer}$ in the $N_{\rm e}=4$ sector.
We then define the following composite states for the full system:
\begin{align}
    \ket{\Phi_{2,m;4,n}} &= \ket{\Psi_{2,m}} 
    \ket{\Psi_{4,n}}, \\
    \ket{\Phi_{4,m;2,n}} &= \ket{\Psi_{4,m}} 
    \ket{\Psi_{2,n}}.
\end{align}

Let \(\ket{\Phi}\) belong to the set \(\{\ket{\Phi_{2,m;4,n}}, \ket{\Phi_{4,m;2,n}}\}_{m,n=1,2,...}\), and let \(\ket{\sigma}\) represent the spin states \(\{\ket{\Uparrow}, \ket{\Downarrow}\}\).
The matrix elements of the high-energy part of the Hamiltonian is determined by the relations
\begin{align}
\langle \sigma_1| \langle \sigma_2| H_{\text{high}} | \sigma'_1 \rangle |\sigma'_2\rangle & = 0, \\
\langle \sigma_1| \langle \sigma_2| H_{\text{high}} |\Phi\rangle & = \langle \Phi| H_{\text{high}} | \sigma'_1 \rangle |\sigma'_2\rangle = 0, \\
\bra{\Phi_{2,m;4,n}} H_{\text{high}} \ket{\Phi_{2,m';4,n'}} &= \delta_{m,m'} \delta_{n,n'} (\langle \Psi_{2,m} | H_{\text{trimer}} | \Psi_{2,m} \rangle + \langle \Psi_{4,n} | H_{\text{trimer}} | \Psi_{4,n} \rangle), \\
\bra{\Phi_{4,m;2,n}} H_{\text{high}} \ket{\Phi_{4,m';2,n'}} &= \delta_{m,m'} \delta_{n,n'} (\langle \Psi_{4,m} | H_{\text{trimer}} | \Psi_{4,m} \rangle + \langle \Psi_{2,n} | H_{\text{trimer}} | \Psi_{2,n} \rangle), \\
\bra{\Phi_{2,m;4,n}} H_{\text{high}} \ket{\Phi_{4,m';2,n'}} &= \bra{\Phi_{4,m;2,n}} H_{\text{high}} \ket{\Phi_{2,m';4,n'}} = 0.
\end{align}

Next, the perturbation term is introduced into the Hamiltonian, with \(t'_2\) ($<0$) taken as the perturbation that couples the two trimer units. The perturbation Hamiltonian is given by
\begin{equation}
H_{\text{pert}} = t'_2 \sum_{\sigma=\uparrow,\downarrow} \left( c_{2,\sigma}^\dagger c_{4,\sigma} + c_{3,\sigma}^\dagger c_{5,\sigma} + \text{H.c.} \right).
\end{equation}

Let ${\cal P}$ be the projection onto the ground states of the unperturbed system such that ${\cal P}|\sigma\rangle |\sigma'\rangle = |\sigma\rangle |\sigma'\rangle$. The effective Hamiltonian in second-order perturbation theory\cite{ye2024kinetic,janani2014low} is obtained as
\begin{equation}
H_{\text{eff}} = - {\cal P} H_{\text{pert}}  \frac{1}{H_{\text{high}} - E_{\text{gs}}} H_{\text{pert}} {\cal P}
\label{Heff}
\end{equation}
where $E_{\text{gs}}$ is the gs 
energy of the unperturbed system. The effective Heisenberg interaction strength, \(J_{\text{eff}}\) is obtained from the difference between the spin-triplet and singlet energies, \(J_{\text{eff}} = E_t - E_s\), where \(E_t\) and \(E_s\) are computed using second-order perturbation theory by solving the Eq. (\ref{Heff}). This leads to the effective exchange Hamiltonian:
\begin{equation}
H_{\text{eff}} = J_{\text{eff}}\hspace{0.1cm} (\mathbf{S}_{\rm L} \cdot \mathbf{S}_{\rm R}),
\end{equation}
where $\mathbf{S}_{\rm L}$ and $\mathbf{S}_{\rm R}$ denote the effective spins of the left and right trimers, respectively. 

Table \ref{x} shows the values of $J_{\text{eff}}/|t'_2|^2$ with $t_2$ for a fixed $U$, $U=1.0$. Clearly, for small \(t'_2\) and large $U$, the ground state 
of the two-trimer system forms a triplet. 
This suggests that, with small \(t'_2\) and large \(U\), the system is expected to transition to a ferrimagnetic phase, which is consistent with the numerical results obtained in the main text. 

\begin{table*}
    \centering
    \setlength{\tabcolsep}{20pt} 
    \renewcommand{\arraystretch}{1.6} 
    \captionsetup{justification=centering} 
    \caption{Dependence of $J_{\text{eff}}/|t'_2|^2$ on $t_2$ for $t_1=-1.0$ and $U=1.0$.}
    \label{tab:my_table}
    \vspace{0.5em} 
    \begin{tabular}{|>{\centering\arraybackslash}p{2cm}|>{\centering\arraybackslash}p{5cm}|} 
        \hline
        $t_2$& $J_{\text{eff}}/|t'_2|^2$\\ 
        \hline 
        $-0.1$ & $-0.155789$ \\ 
        $-0.2$ & $-0.161683$ \\ 
        $-0.3$ & $-0.172489$ \\ 
        $-0.4$ & $-0.19005$ \\ 
        \hline
    \end{tabular}
    \label{x}
\end{table*}
\bibliography{sup_ref}